\documentclass[superscriptaddress,reprint,amsmath,amssymb,aps,pra,floatfix]{revtex4-2}

\usepackage{graphicx}
\usepackage{dcolumn}
\usepackage{bm}
\usepackage{hyperref}
\usepackage{xcolor}

\begin{document}

\title{
Dynamical phase diagram of a one dimensional Bose gas in a box with a tunable weak-link:
from Bose-Josephson oscillations to shock waves
}

\author{Abhik Kumar Saha}
\affiliation{School of Physical Sciences, Indian Association for the Cultivation of Science, Jadavpur, Kolkata 700032, India.}

\author{Romain Dubessy}
\email{romain.dubessy@univ-paris13.fr}
\affiliation{Université Sorbonne Paris Nord, Laboratoire de Physique des Lasers, CNRS, UMR 7538, F‐93430, Villetaneuse, France}

\date{\today}

\begin{abstract}
We study the dynamics of one-dimensional bosons trapped in a box potential, in the presence of a barrier creating a tunable weak-link, thus realizing a one dimensional Bose-Josephson junction.
By varying the initial population imbalance and the barrier height we evidence different dynamical regimes.
In particular we show that at large barriers a two mode model captures accurately the dynamics, while for low barriers the dynamics involves dispersive shock waves and solitons.
We study a quench protocol that can be readily implemented in experiments and show that self-trapping resonances can occur. This phenomenon can be understood qualitatively within the two-mode model.
\end{abstract}


\maketitle
\clearpage
\section{Introduction}
The difference between classical and quantum dynamics becomes strikingly evident when two macroscopic quantum objects are coupled by a weak link. For example, the system consisting of two superconductors separated by a thin insulating layer has attracted much attention due to the discovery of the Josephson effect~\cite{Josephson1962}: a direct current can flow in this setup, even without applying any external voltage. Later the first experimental observation of the Josephson effect~\cite{Anderson1963}, has opened the way to many applications, including its generalization to other setups~\cite{Likharev1979,Barone1982}.

Although the theory of the Josephson junction was originally developed in the context of superconductivity, it can be applied as well to describe two weakly coupled Bose-Einstein condensates (BEC)~\cite{Smerzi1997,Milburn1997,Zapata1998,Raghavan1999}, by using for example a double-well potential, thus realising an atomic Bose-Josephson junction (BJJ). Due to two-body interaction between atoms the BJJ exhibits new dynamical regimes such as macroscopic quantum self-trapping (ST)~\cite{Raghavan1999,Raghavan1999b}, not present in the superconductor Josephson junction. This new effect, as well as the observation~\cite{Cataliotti2001} of Josephson oscillations (JO), have been observed in a single BJJ~\cite{Albiez2005,Levy2007}. This has raised a lot of interest in the study of BJJ and ongoing theoretical and experimental studies deal with dissipative ~\cite{Smerzi1997,Meier2001,LeBlanc2011,Pigneur2018,Saha2020} and non-dissipative oscillations~\cite{Smerzi2003,Spagnolli2017,MartinezGaraot2018}, supercurrent dynamics in ring shaped condensates~\cite{Ramanathan2011,Ryu2013,Polo2019}, current phase relation of atomic BJJ~\cite{Eckel2014,Jendrzejewski2014}, quantum transport~\cite{Krinner2017,Gamayun2021}, as well as their counterparts with fermionic superfluid atomic samples~\cite{Valtolina2015,Burchianti2018,Zaccanti2019,Luick2020,Xhani2020a,Kwon2020}.

When studying the atomic Josephson junction several factors must be taken into account among which the geometry and the effective dimension are of particular importance. Previous works have dealt with the double-well geometry, in which two elongated gases are side by side~\cite{Shin2004,Schumm2005,Mennemann2020}, or the ring geometry in which the gases are coupled head to toe~\cite{Bidasyuk2016}. The former geometry is adapted for matter wave interferometry~\cite{Shin2004,Schumm2005,Gati2006}, while the latter enables the realization of atomic circuits~\cite{Ryu2013,Edwards2013,Eckel2014,Amico2017,Ryu2020}. The effective dimension is crucial to determine the excitations involved in the dynamics~\cite{Abad2015,Saha2019}: in three dimensions vortex lines decaying into vortex rings~\cite{Piazza2011}, in two dimensions point-like vortices carrying phase-slips~\cite{Singh2020,Griffin2020a} and in one dimension (1D), for which no vortex can exist, solitons and dispersive shock waves (DSW)~\cite{Polo2019,Simmons2020,Dubessy2021}. Furthermore, in all dimensions the excess energy can be dissipated into a phonon bath, allowing for a relaxation of the JO. In ultracold atoms experiments the effective dimension is controlled by the transverse confinement and can be tuned~\cite{Mewes1996,Gorlitz2001,Salasnich2004}.

In this work we consider a zero temperature 1D Bose gas confined in a box potential with a tunable central barrier, thus defining two (left and right) reservoirs connected through a tunable weak link. In this geometry each reservoir contains many excitation modes that contribute to the dynamics, in contrast to the simple two-mode model picture~\cite{Raghavan1999}. As we will show below this geometry is interesting because it evidences the interplay between shock wave dynamics and Josephson physics. In particular we find a clear unified framework to describe the whole dynamical phase diagram, from the weak coupling regime, where the two-mode model is valid~\cite{LeBlanc2011}, to the large coupling regime, where shock wave emerge. Furthermore we observe that soliton nucleation at the weak-link can induce a fast damping of the density oscillations. We consider quench protocols that can be readily implemented in experiments and show how the quench speed affects the dynamics by inducing self-trapping resonances.

This paper is organised as follows: in Sec.~\ref{sec:model} we describe the model we study and the tools we use, in Sec.~\ref{sec:dynamical} we report on a thorough study of the dynamical phase diagram of the 1D Bose gas. We then discuss in Sec.~\ref{sec:quench} how the quench protocol modifies the dynamical phase diagram and finally discuss how our results open new perspectives.

\section{\label{sec:model}Model and quench protocol}
We consider $N$ weakly interacting bosons of mass $m$ with repulsive contact interaction on a 1D ring of circumference $L$, described at zero temperature by the mean-field Gross-Pitaevskii equation (GPE):
\begin{eqnarray}
i\hbar\frac{\partial \psi}{\partial t}
=\left(-\frac{\hbar^2}{2m}\frac{\partial^2}{\partial x^2}+V_{\rm ext}(x,t) +g_{1D}N|\psi|^2\right)\psi
\label{GPE}
\end{eqnarray}
where $\psi(x,t)$ is the condensate wave function, normalized to unity $\int_0^L dx\,|\psi|^2=1$, $g_{1D}=2\hbar\omega_{\perp}a_{s}$ is the 1D interaction strength~\cite{Olshanii1998}, where $\omega_{\perp}$ is the radial confinement frequency and $a_{s}$ is the three-dimensional $s$-wave scattering length, and $V_{\rm ext}(x,t)$ is the external trapping potential.

\begin{figure}[b]
\centering
\includegraphics[width=8cm]{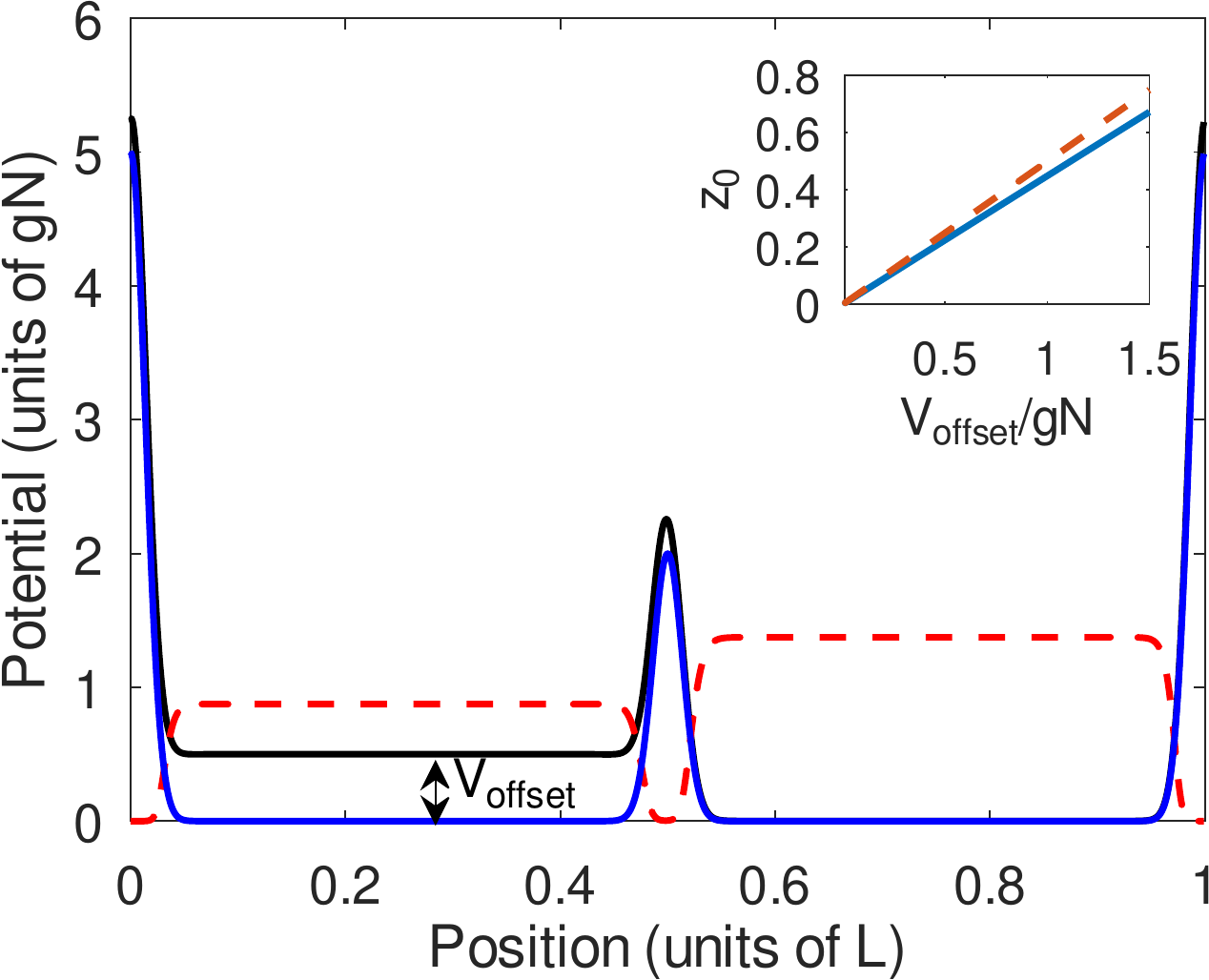} 
\caption{\label{fig:quench}(color online)
Setup and quench considered in the present work. Initially a 1D Bose gas is prepared in two reservoirs with a population imbalance, using an energy offset (double arrow) as shown with the solid black potential curve. At $t=0$ the energy offset is removed resulting in a new potential landscape show as the solid blue curve. The initial atomic density distribution is sketched by the dashed red curve. The inset shows the initial population imbalance $z_0$ as a function of $V_{\rm offset}$ obtained with this method (solid blue line), compared to the naive expectation $z_0=V_{\rm offset}/(2gN)$ (dashed red line).
}
\end{figure}

We consider an external trap potential made by a combination of two Gaussian barriers:
\[
V_{\rm trap}(x,t)=V_{0}\exp\left[-\frac{x^2}{\sigma^2}\right]+V(t)\exp\left[-\frac{(x-\frac{L}{2})^2}{\sigma^2}\right]
\]
located at $x=0$ and $x=L/2$. Because of the ring geometry and periodic boundary conditions this results in a effective box-like potential for large $V_0$ with a tunable barrier $V(t)$ separating two reservoirs see Fig.~\ref{fig:quench}.
We create the initial population imbalance between the two reservoirs by using an auxiliary potential:
\[
V_{\rm imb}(x)=\frac{V_{\rm offset}}{2}\left[\tanh{\left[\frac{x}{\sigma}\right]}+\tanh{\left[\frac{x-L/2}{\sigma}\right]}\right]
\]
where $V_{\rm offset}$ controls the initial population imbalance, as shown in the inset of Fig.~\ref{fig:quench}.

In this setup we expect that the relevant energy scale is fixed by the bare chemical potential $\mu_0=g_{1D}n_0$, where $n_0=N/L$ is the typical density, with a healing length $\xi=\hbar/\sqrt{2mg_{1D}n_0}$. We consider a large $V_{0}=5\mu_{0}$ barrier to create the box with relatively narrow $\sigma=4\xi$ width.

To solve Eq.~\eqref{GPE} we use its dimensionless form, in which lengths  are scaled by $L$, time by $T=mL^2/\hbar$ and interaction strength by $\hbar^2/(mL)$. Using these units Eq.~\eqref{GPE} becomes
\begin{equation}
i\frac{\partial\psi}{\partial t}=\left(-\frac{1}{2}\frac{\partial^2}{\partial x^2}+V_{\rm ext}(x,t) +gN|\psi|^2\right)\psi.
\label{dGPE}
\end{equation}
To numerically solve Eq.~\eqref{dGPE} we use a spectral method relying on fast Fourier transforms to evaluate exactly the kinetic energy term~\cite{Blakie2008}, with a regular grid of 256 points and a large dimensionless non linear parameter $gN=20000$, well within the mean-field regime.
The time integration is performed using a standard fourth order Runge-Kutta scheme with a typical time step of $\delta t=2\times10^{-5}$~\footnote{We use MATLAB ode45 integrator and have checked that relative errors on total atom number and energy remain in the $10^{-10}$ to $10^{-6}$ range for all simulations.}.
We note that the choice of this spectral method imposes the use of periodic boundary conditions, hence the ring geometry, but the large barrier at $x=0$ transforms it to a box potential. The interplay between box and ring geometry has been already studied in Ref.~\cite{Bidasyuk2016}.

To initialize the system we use imaginary time propagation~\cite{Antoine2007,Barenghi2016} in Eq.~\eqref{dGPE}, in the presence of both the static initial trap and imbalance potentials: $V_{\rm ext}(x)=V_{\rm trap}(x,0)+V_{\rm imb}(x)$. Once the evolution has converged to the groundstate~\footnote{During imaginary time propagation we monitor the chemical potential at each step and stop when relative changes are below the $10^{-12}$ level.} we abruptly remove the imbalance potential to initiate the dynamics.
We then explore two situations: $V(t)$ is either kept constant, see section~\ref{sec:dynamical}, or quenched in a time $\tau$ to a lower value, see section~\ref{sec:quench}.
For each case we study how the central barrier strength $V_1$ affects the dynamics. 

\section{\label{sec:dynamical}Dynamical regimes for a one dimensional Bose gas}
In this section we consider a static barrier $V(t)=V_1$ and study  the effect of $V_{\rm offset}$, that is the initial population imbalance between the reservoirs. After the preparation, we quench the system by removing abruptly the imbalance potential at $t=0$ and study the dynamics.

To analyze the dynamics we measure the time dependent density $n(x,t)=|\psi(x,t)|^2$, the total current per particle~\footnote{We use here the expectation value of the momentum operator.} $J(t)=\frac{\hbar}{imL}\int_0^Ldx\,\psi^*\frac{\partial\psi}{\partial x}$, and the population imbalance between the two reservoir: $z(t)=\int_{L/2}^L dx\,|\psi|^2-\int_0^{L/2}dx\,|\psi|^2$. As expected from previous studies of the BJJ, we find mainly three different regimes for the population imbalance dynamics~\cite{Xhani2020}: oscillations, self-trapping and damped oscillations. However we also uncover particular regimes that are unique to the 1D geometry, involving dispersive shock waves and solitons.

\begin{figure}[t]
\centering
\includegraphics[height=4.7cm,width=8.6cm]{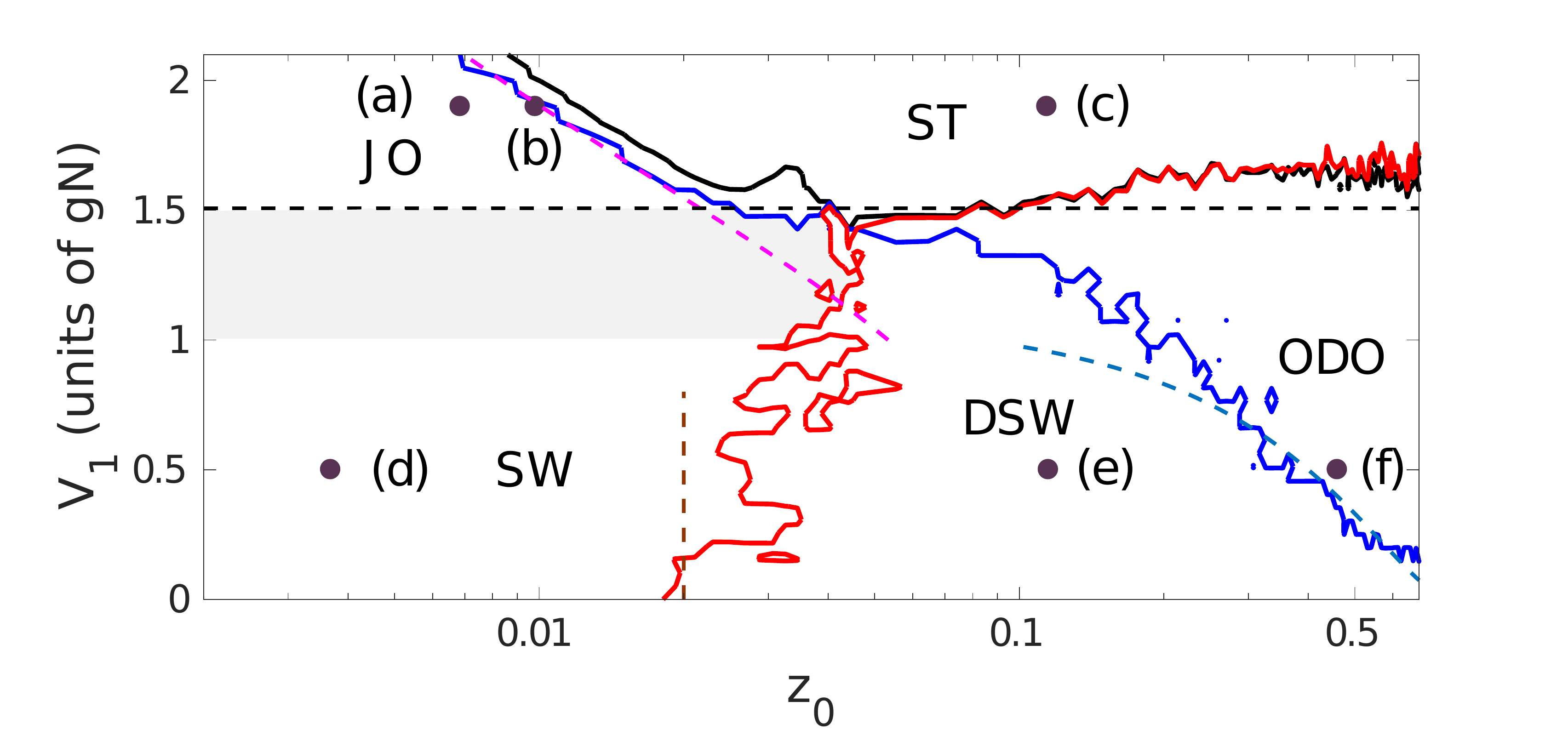} 
\caption{\label{fig:PSD}
(color online) Dynamical phase diagram of a one dimensional Bose gas in a box, varying the central barrier strength ($V_1$) and the initial imbalance $z_0$ (controlled by $V_{\rm offset}$), based on the three criteria described in the text, defining four main regimes: self trapping (ST), dispersive shock wave (DSW), over damped oscillations (ODO) and regular oscillations: with a smooth crossover (grey shaded area) from Josephson oscillations (JO) to shock wave (SW). The dashed lines indicate analytical results. Note the logarithmic scale on the horizontal axis. The labels (a-f) refer to the examples shown in Fig.~\ref{fig:regimes} and Table~\ref{tab:parameters}.
}
\end{figure}

In order to classify simply the different regimes, we will use three quantitative figure of merit, as shown in Fig.~\ref{fig:PSD}. The first two are based on the power spectrum density of $z(t)$, defined as: $C(\omega)=|\hat{z}(\omega)|^2$, where $\hat{z}(\omega)$ is the Fourier transform of $z(t)$. As is well known in signal analysis, $C(\omega)$ is a measure of the power distribution among frequencies in a signal. To determine if the system is time dependent we compute the relative weight of the zero frequency term in to the total power of the signal: $C_0=C(0)/\int d\omega\,C(\omega)\in[0,1]$. By computing $C_0$ for various $V_1$ and $V_{\rm offset}$ we observe that it defines well separated regions of high (close to 1) and low (close to 0) values, with sharp boundaries. We choose arbitrarily the value $C_0=0.95$ to distinguish the different regimes (solid black line in Fig.~\ref{fig:PSD}): for $C_0>0.95$ the dynamics is ``frozen" while for $C_0<0.95$ the time evolution is dominant. Then, to distinguish between the different dynamical regimes, we compute the frequency $\omega_{\rm max}$ at which the maximum of $C(\omega)$ occurs. In order to define a criteria, we compare it to a typical frequency, related to sound propagation in the system: $\omega_s=\pi\times c/L$, where $c$ is the initial speed of sound in the left reservoir. We consider that the system is mainly oscillating when $\omega_{\rm max}>\omega_s/10$ (solid blue line in Fig.~\ref{fig:PSD}). Finally to decide if the dynamics is relaxing towards a steady state we evaluate the damping time by computing the half amplitude decay time: $t_{\rm half}=\textrm{max}\,\{t,~\textrm{such that}~|z(t)|>|z_0|/2\}$, where $z_0$ is the initial imbalance, and consider that the dissipation is small when $t_{\rm half}>0.35\times mL^2/\hbar$ (solid red line in Fig.~\ref{fig:PSD}).

Figure~\ref{fig:PSD} shows the different dynamical regimes of the 1D Bose gas when the central barrier height $V_1$ and the initial population imbalance $z_0$ are varied. We identify four regimes: frozen dynamics, corresponding to the self-trapping (ST) regime, for $C_0>0.95$, or equivalently $\omega_{\rm max}<\omega_s/10$ and $t_{\rm half}>0.35\times mL^2/\hbar$; regular oscillations
for $C_0<0.95$, $\omega_{\rm max}>\omega_s/10$, $t_{\rm half}>0.35\times mL^2/\hbar$; damped dynamics, corresponding to dispersive shock wave (DSW), for $C_0<0.95$, $\omega_{\rm max}>\omega_s/10$ and $t_{\rm half}<0.35\times mL^2/\hbar$; over damped dynamics, corresponding to over damped oscillations (ODO), for $C_0<0.95$, $\omega_{\rm max}<\omega_s/10$ and $t_{\rm half}<0.35\times mL^2/\hbar$.
As discussed below, 
at large barriers we find results very close to the regular two-mode model of the BJJ~\cite{LeBlanc2011}, while at low barriers we find that the dynamics are mediated by propagating shock waves. Therefore, in the regular oscillations regime, we distinguish the (JO) and shock wave (SW) regimes: the light grey area indicates approximately the transition region from JO to SW. The SW regime is uniquely identified by the fact that the density imbalance oscillations occurs exactly at the sound frequency. 

We note that the three criterion are in a reasonable agreement to define the ST regime, which indicates that the arbitrary thresholds we choose are self consistent. In order to confirm this interpretation of the dynamical phase diagram we show example trajectories in Fig.~\ref{fig:regimes}, see below.

\begin{figure*}[t]
\centering
\includegraphics[width=17.8cm]{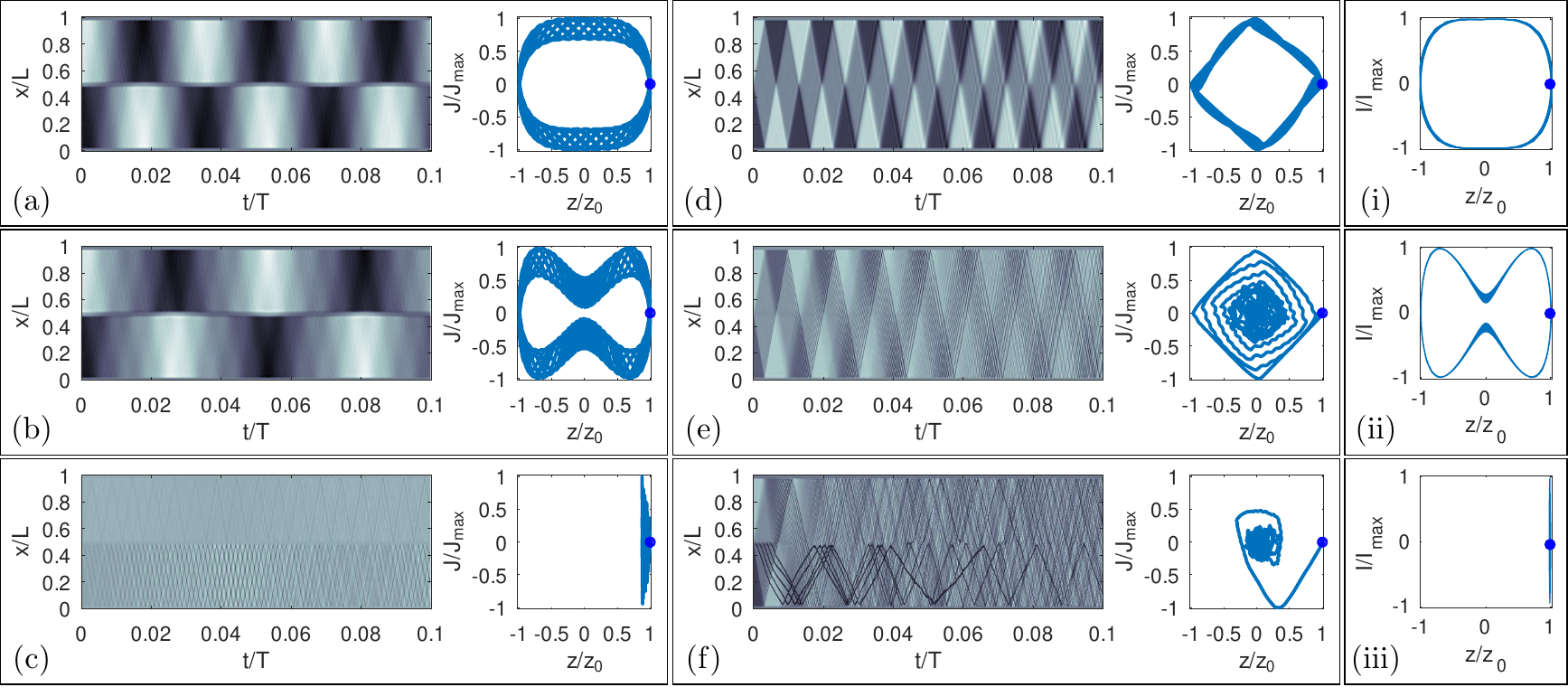}
\caption{\label{fig:regimes}(color online)
Detail of the density fluctuations dynamics in different regimes (colormap), corresponding to the parameters of Table~\ref{tab:parameters} and the phase portrait showing the normalized current ($J/J_{\rm max}$) versus normalized population imbalance ($z/z_0$).
(a) JO, (b) transition to ST, (c) ST regime, (d) SW, (e) DSW and (f) ODO. (i), (ii) and (iii) phase portrait of the two mode model for the same parameters as in (a), (b) and (c) respectively. In all phase portrait the blue disk indicate the initial condition ($z=z_0$ and $J=0$). See text for details.
}
\end{figure*}

In order to gain more insight into this dynamical phase diagram we show in Fig.~\ref{fig:regimes} a few example cases corresponding to the labels (a-f) in Fig.~\ref{fig:PSD} and simulation parameters detailed in Table~\ref{tab:parameters}. We have found that the best way of studying the various dynamical regimes is to plot the density fluctuations: $\delta n(x,t)=n(x,t)-\frac{1}{T}\int_0^T dt\,n(x,t)$ and the current $J(t)$ versus imbalance $z(t)$ trajectory. For the sake of clarity we display only the early time dynamics for density fluctuations (up to $t=0.1\times mL^2/\hbar$). Studying the density fluctuations allows to remove the background density variations imposed by the barriers and thus focus on the excitations propagating through the system. Plotting the $J(t)$ versus $z(t)$ trajectory allows a direct comparison with other realizations of Josephson junctions and in particular emphasizes the specificity of the 1D case. In order to compare different regimes we use a normalized plot, where the current is normalized to its maximum value and the imbalance is normalized to its initial value.

\begin{table}[b]
\caption{\label{tab:parameters}Parameters used in Fig.~\ref{fig:regimes} to illustrate the different dynamical regimes.
}
\begin{ruledtabular}
\begin{tabular}{ccccl}
& $V_1/gN$ & $V_{\rm offset}/gN$ & $z_0$ & Regime\\
\colrule
(a) & 1.9 & 0.015 & 0.007 & Josephson oscillation (JO)\\
(b) & 1.9 & 0.0215 & 0.010 & JO to ST transition point\\
(c) & 1.9 & 0.25 & 0.114 & self-trapping (ST)\\
(d) & 0.5 & 0.008 & 0.004 & shock wave (SW)\\
(e) & 0.5 & 0.25 & 0.115 & dispersive shock wave (DSW)\\
(f) & 0.5 & 1 & 0.460 & over-damped oscillations (ODO)\\
\end{tabular}
\end{ruledtabular}
\end{table}

Fig.~\ref{fig:regimes}(a) shows a regime very similar to the usual two-mode model of the BJJ, demonstrating that for small initial imbalance and large barrier height the dynamics involves few modes, exhibiting a quasi circular $J(t)$ vs $z(t)$ trajectory.

Fig.~\ref{fig:regimes}(b) shows the intermediate regime between JO and ST, also reminiscent of the simple two-mode model.

Fig.~\ref{fig:regimes}(c) shows the ST regime, where the dynamics is quasi frozen and the imbalance remains very close to its initial value. The small propagating density oscillations are induced by the initial quench, however the excitations in the two reservoirs remain decoupled.

Fig.~\ref{fig:regimes}(d) shows the SW appearing at low barrier and low initial imbalance, for which regular oscillations occur with extremely small damping. The density oscillations show almost piece-wise constant profiles with sharp fronts propagating at a well defined velocity. This can be understood in the framework of the recent prediction of universal shock wave dynamics in quenched 1D Bose gases trapped in a box~\cite{Dubessy2021}. The quench protocol studied here results also in two counter-propagating dispersive shock waves that bounce on the box boundaries.

Fig.~\ref{fig:regimes}(e) shows damped oscillations in a regime dominated by dispersive shock wave propagation~\cite{Dubessy2021}: as the fronts propagate the rarefaction front broadens while the shock front dissolves into a soliton train, as described by Whitham's modulation theory~\cite{El2016a}. In this regime the effect of the barrier is small.

For both Fig.~\ref{fig:regimes}(d) and (e) the oscillation is sustained by a fully nonlinear propagating density shock wave, resulting in the peculiar diamond shape of the $J(t)$ versus $z(t)$ trajectory, as opposed to the usual two mode BJJ dynamics of (a) and (b).

Fig.~\ref{fig:regimes}(f) shows the over damped regime, for which the $J(t)$ versus $z(t)$ trajectory quickly relaxes to the origin. A close inspection of the density fluctuations shows that this quick damping is associated to the nucleation of individual solitons at the central barrier, on the density depleted side, that remain confined in their reservoir. This soliton nucleation process occurs when the local density at the barrier vanishes~\cite{Hakim1997}, enabled by the large density fluctuations at increasing $z_0$.

This study evidences the peculiarity of the 1D Bose gas dynamics in contrast with the recent work of~\cite{Xhani2020} that investigated the 3D BJJ. As is well known in 1D physics, there are fewer available decay channels resulting in long lived excitations as dispersive shock waves for example~\cite{Simmons2020,Dubessy2021}, that sustain regular oscillations. This can be seen as a consequence of the underlying integrability of the Lieb-Liniger model. As was shown in~\cite{Dubessy2021} the DSW results in a dephasing of the total current and hence, in the context of BJJ, gives a damped oscillation. This damping is therefore not directly related to the effect of the central barrier.
However the barrier is crucial to explain the appearance of the over damped regime, associated to spontaneous soliton nucleation at the barrier, a phenomenon similar to the phase slip mechanism observed in ref.~\cite{Polo2019}.

As shown in Fig.~\ref{fig:PSD} the transition from SW to DSW is almost a vertical line, meaning that it depends mainly on the initial imbalance. This can be understood be considering the dephasing time of the underlying shock wave~\cite{Dubessy2021}. Indeed the shock fronts broaden due to the difference between the highest $c_{\rm max}$ and the lowest $c_{\rm min}$ velocities. We may expect that the oscillation reaches its half amplitude at a time $\tau=L/(c_{\rm max}-c_{\rm min})$. Using then very simple estimates, see appendix \ref{app:sound}, we find: $c_{\rm max}-c_{\rm min}\simeq\sqrt{gN/L}z_0$, and therefore $z_0=L/(\tau\sqrt{gN/L})$. The brown vertical dashed line in Fig.~\ref{fig:PSD} corresponds to a time $\tau=0.35$, and therefore a initial imbalance of: $z_0\simeq0.02$. This very simple estimate is already in good agreement with the simulations, small corrections are expected due to the effect of the box boundaries. We find convenient to distinguish between the SW and DSW regimes based on this arbitrary criterion but we note that there is in fact a smooth and continuous transition from weakly dispersive shock waves to strongly dispersive shock waves as $z_0$ increases.

Similarly one can use modulation theory to estimate when independent solitons can be nucleated at the barrier. In particular the lowest density $n_b$ can be written as a function of the population imbalance, see appendix \ref{app:sound} for details. Then, using local density approximation, the critical barrier for soliton nucleation is: $V_1^s\simeq gn_b$, see Eq.~\eqref{eqn:V1s}, corresponding to the blue dashed line in Fig.~\ref{fig:PSD}, that captures the DSW to ODO transition up to $V_1\simeq gN$.

We note that for large barriers $V_1>1.5\times gN$, the dynamics is very similar to the canonical two-mode model predictions for the BJJ. To compute the relevant parameters for the two-mode model we follow the method of~\cite{Raghavan1999,Xhani2020}. Briefly we derive the usual BJJ equations:
\begin{subequations}
\begin{eqnarray}
\dot{z}&=&-2|K|\sqrt{1-z^2}\sin{\theta},\\
\dot{\theta}&=&\left[U+2|K|\frac{\cos{\theta}}{\sqrt{1-z^2}}\right]z,
\end{eqnarray}
\label{eqn:tmmstd}
\end{subequations}
where the tunneling rate $|K|$ and the interaction energy $U$ can be related to the microscopic parameters of the GPE, see appendix~\ref{app:tmm} for details.
Equation~\eqref{eqn:tmmstd} allows for example to predict the critical imbalance at which the self-trapping transition occurs~\cite{Raghavan1999}, for a given barrier strength $V_1$. The agreement with the full GPE simulation is remarkable, as shown by the dashed magenta line in Fig.~\ref{fig:PSD}.
Using Eq.~\eqref{eqn:tmmstd} one can also predict the current versus imbalance phase portrait of the two-mode model. To define the quantity corresponding to the current $J(t)$, we follow the standard definition in superconductor Josephson junction~\cite{Barone1982,Tinkham1996}, for which the current is given by the rate of change of the population imbalance: $I(t)=\dot{z}(t)$.

Figure~\ref{fig:regimes}(i-iii) show a direct comparison of the current versus imbalance computed within the two-mode model, with the full GPE result, for the same parameters as in Fig.~\ref{fig:regimes}(a-c).
We observe that the shape of the trajectories are correctly captured within the two-mode approximation.
We conclude that our protocol is able to simulate the two-mode Josephson physics at large barriers and allow to study the interplay with shock-wave physics at smaller barriers.

\begin{figure}[ht]
\centering
\includegraphics[width=8cm]{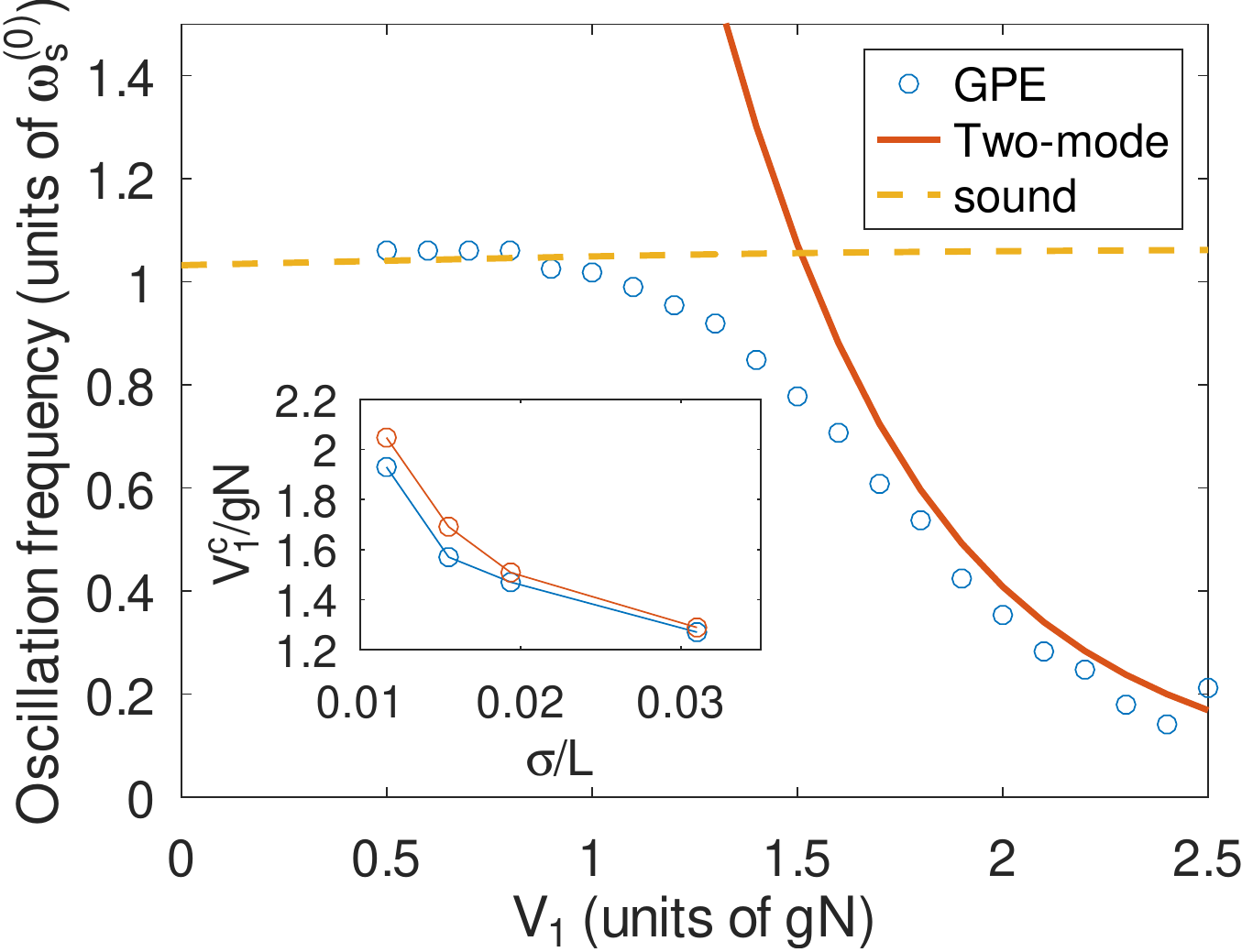}
\caption{\label{fig:cross}(color online)
Main oscillation frequency of the current $J(t)$, normalized to the bare sound frequency $\omega_s^{(0)}=\pi\sqrt{gN}/L$, from the full GP solution for initial imbalance $z_0=0.004$ (blue circles), the two-mode model prediction of Eq.~\eqref{eqn:freqTM} (solid red curve) and the sound frequency (dashed yellow line). The inset shows the critical barrier height $V_1^c$ as a function of the barrier width, see text for details.}
\end{figure}

More precisely we can study quantitatively this crossover by monitoring the oscillation frequency at small imbalance as a function of the central barrier height $V_1$, see Fig.~\ref{fig:cross}, for which the two-mode model predicts:
\begin{equation}
\omega_{TM}=2|K|\sqrt{1+\frac{U}{2|K|}}.
\label{eqn:freqTM}
\end{equation}
As $\omega_{TM}$ diverges when $V_1$ decreases, one can define a critical barrier $V_1^c$ such that $\omega_{TM}(V_1^c)=\omega_s$, which provides a natural upper limit associated to sound mediated transport.
Interestingly we have found that $V_1^c$ is always very close to the lowest boundary of the self-trapping regime, as shown in the inset of Fig.~\ref{fig:cross} for various central barrier width. This allows us to add a horizontal black dashed line in the dynamical phase diagram of Fig.~\ref{fig:PSD}, above which the two-mode model is valid, for all imbalances. We note that for small imbalance the SW regime is reached when $V_1\leq\mu$, which corresponds to direct transport above the barrier, without tunneling effects.

Our work thus provides a comprehensive study of the dynamical phase diagram of a 1D Bose gas in a box with a tunable barrier, evidencing the interplay of Josephson oscillations and shock-waves dynamics.

\section{\label{sec:quench}Dynamics after a barrier quench}

\begin{figure}[b]
\centering
\includegraphics[width=8cm]{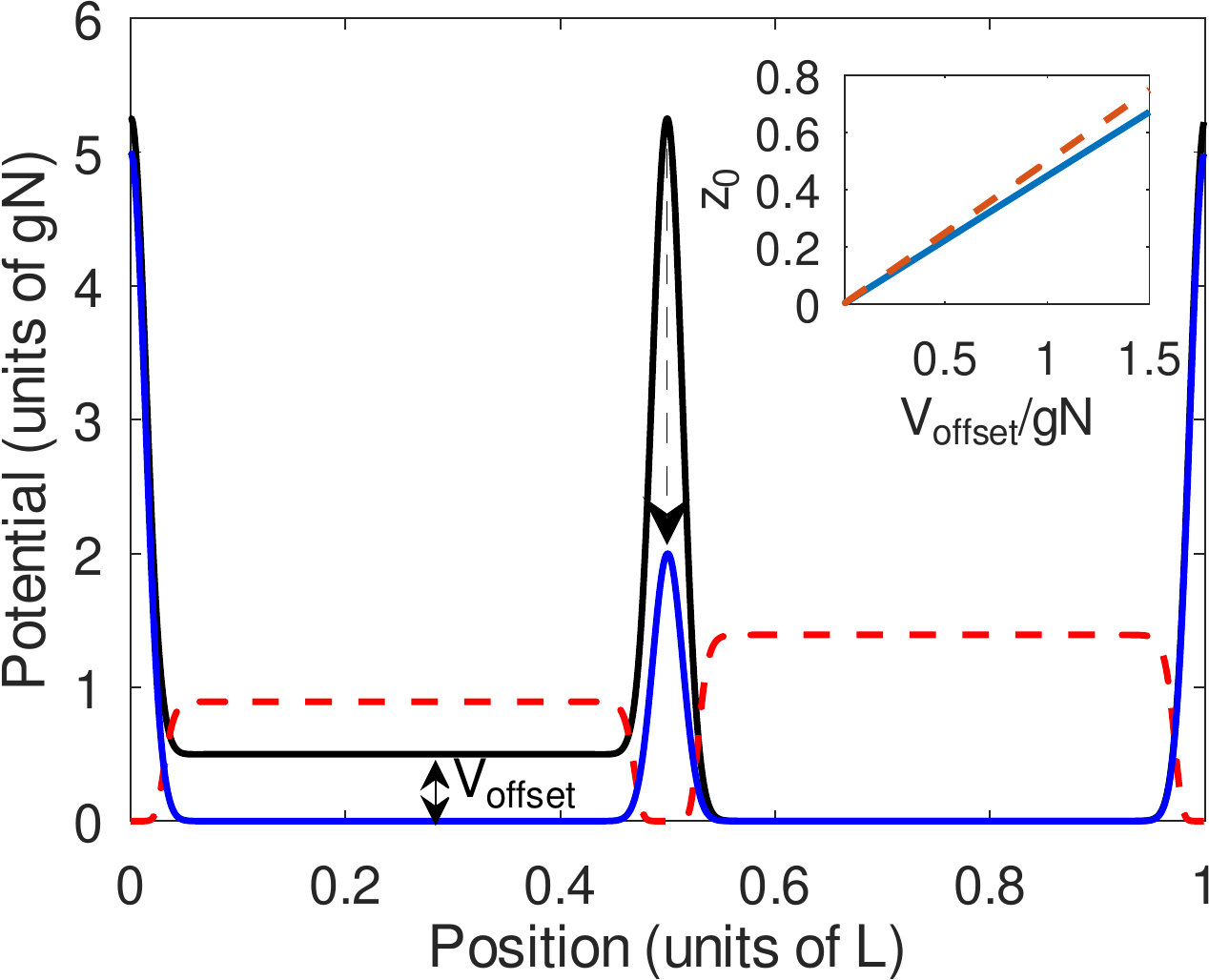} 
\caption{\label{fig:quench2}(color online)
Modification of the quench protocol: the system is initialized with a large central barrier, quenched to its final value in a time $\tau$, as sketched by the dashed arrow. As in Fig.~\ref{fig:quench} the solid black line is the initial potential, the blue line is the post quench potential and the dashed red line is the initial density profile. The inset shows $z_0$ as a function of $V_{\rm offset}$ obtained with this method.
}
\end{figure}

\begin{figure}[b]
\centering
\includegraphics[width=8.6cm]{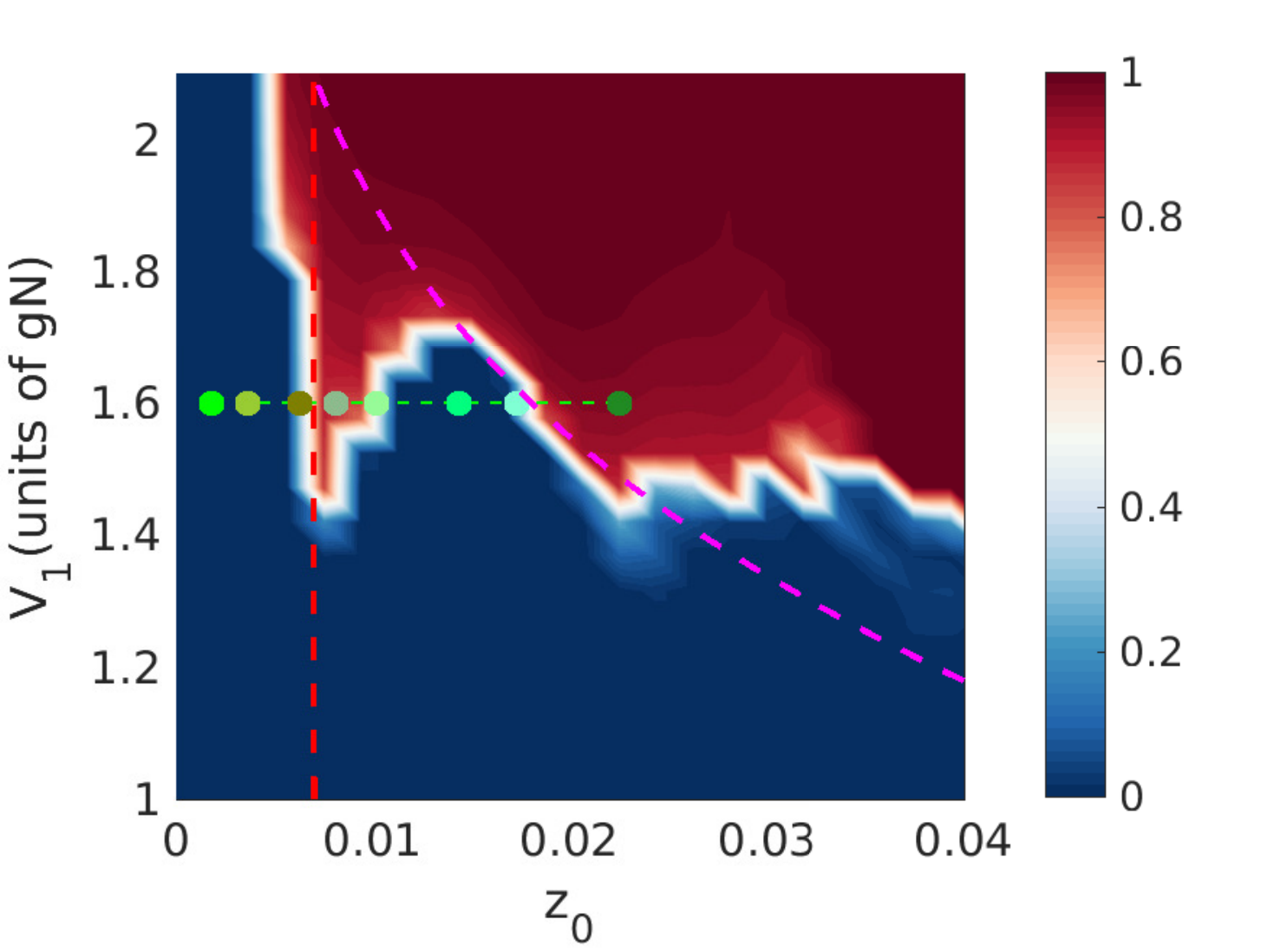} 
\caption{\label{fig:tau:0.01}(color online)
Normalized power spectrum density at zero frequency as a function of central barrier height $V_{1}$ and initial imbalance $z_0$ for a quench of $\tau=0.01$. The magenta dashed line indicates the prediction of the two-mode model for the self-trapping transition. The red vertical dashed line corresponds to a self-trapping resonance, as discussed in the text. The green line and markers correspond to the example trajectories shown in appendix~\ref{app:tau}, see Fig.~\ref{fig:extraTraj}.
}
\end{figure}

In this section, we consider an additional quench of the barrier strength: 
\begin{equation}
V(t)=
\begin{cases}
V_{0}+(V_{1}-V_{0})\frac{t}{\tau}  & t<\tau, \\
V_{1} & t\geq\tau.
\end{cases}
\label{barrierquench}
\end{equation}
The population imbalance is initially prepared in two decoupled reservoirs at large barrier using imaginary time propagation as in section~\ref{sec:dynamical}. Then, at $t=0$ the imbalance potential is removed and Eq.~\eqref{dGPE} is solved while the central barrier height is linearly decreased to its final value according to Eq.~\eqref{barrierquench}. This quench protocol is relevant to experiments as it ensures a clean preparation of two independent reservoirs by initially suppressing the coupling through the central barrier and restoring it in a controlled way, as shown in Fig.~\ref{fig:quench2}.
As in section~\ref{sec:dynamical} we investigate the tunneling dynamics between the two reservoirs as a function of the initial imbalance and the final barrier strength $V_1$.
The main, surprising, difference we find is an inhibition of the Josephson oscillations thus expanding the self-trapping regime: we call this a self-trapping resonance (depending on $z_0$) and show hereafter that it can be predicted from the two-mode model equations~\eqref{eqn:tmmstd}.
We evidence this behavior by focusing on the $C_0$ value, as shown in Fig.~\ref{fig:tau:0.01}.

We find that at low initial imbalance the self-trapping region extends to unusual low barrier values, well below the two mode model prediction. To confirm this unexpected behavior we study a few trajectories, shown in Fig.~\ref{fig:extraTraj} and corresponding to increasing values of initial imbalance at fixed final barrier $V_1=1.6\times gN$. This confirms that the system undergoes a first transition from a Josephson oscillation regime to self trapping, then reverts to an oscillating behavior before reaching the usual self-trapping regime.
Due to the extra energy provided by the quench the maximum imbalance can be larger than $z_0$ and therefore we normalize the phase portrait by $z_{\rm max}$. However we observe that the shape of the trajectories remains comparable to the two mode model predictions.
We have tested several values of the quench time and found qualitatively the same behavior for a wide range of quenching times from $\tau=0.005$ to $\tau=0.1$.

This self-trapping resonance can be explained qualitatively within the two mode model~\eqref{eqn:tmmstd}.
The crucial point here is that the initial barrier is high so that the system is always in the self trapping at $t=0$. Then for $t<\tau$ the barrier is lowered and the self-trapping persists up to a given barrier height. During this time the imbalance remains approximately constant $z(t)\simeq z_0$ but the phase evolves and using the fact that $U\gg|K|$ we find: $\theta(t)=Uz_0 t$. When the dynamics is still frozen at $t=\tau$, this accumulated phase difference modifies the critical point of the self-trapping regime. The critical barrier is then minimal (see Eq.~\eqref{eqn:critical}) at the resonance condition $\theta(\tau)=n \pi$, with $n\in\mathbb{N}^*$, that is for $z_0=n\pi/(U\tau)$. For our parameters this gives $z_{0,\rm res}=0.007$ in good agreement with the observed resonance.

To test this explanation we repeat the same procedure for various values of $\tau$. We find that the number of resonances increase with $\tau$ and that their position is in good agreement with our simple prediction, see appendix~\ref{app:tau} for details.

\section{Conclusion}
In this work we show that several quantitative criteria can be combined to uniquely determine the dynamical diagram of the one dimensional Bose gas and identify the different regimes. We uncover the interplay between Bose-Josephson oscillations and shock wave propagation and show that the same experimental protocol can be used to produce both.
Importantly our analysis rely only on a measure of the time dependent density oscillations, which is routinely achieved in ultra-cold atom experiments.
Furthermore the realization of a one dimensional Bose gas confined in a box potential, including a tunable weak-link is within reach~\cite{Mohammadamin2019}.

We also demonstrate that the phase portrait (current versus imbalance) is an appropriate tool to compare the numerical simulations to simple analytical models and that it is sufficient to identify the different regimes. In particular, despite the intrinsic multi-mode dynamics, we evidence that the main dynamical features are well captured by the two-mode model, for sufficiently large barriers.

It is worth emphasizing that the dispersive shock wave dynamics observed in this work is expected to be universal with respect to the interaction strength, upon a proper rescaling of the dynamical quantities~\cite{Dubessy2021}. It would be interesting to investigate this assumption using exact methods at large interactions strength or a hydrodynamic description at arbitrary interactions.

\begin{acknowledgments}
A.K.S is grateful to Bimalendu Deb for initial discussion of manuscript.
LPL is UMR 7538 of CNRS and Sorbonne Paris Nord University.
\end{acknowledgments}

\appendix

\section{\label{app:sound}Estimates for dispersive shock waves}
We provide here simple estimates to compute the relevant speed of sound in the system. We will assume that finite size effects are small and in particular neglect the bend of the wavefunction near the box boundaries, that occurs on a scale of the healing length $\xi\sim1/\sqrt{2gn_0}$, where $n_0\equiv N/L$ is the average density. We recall that the speed of sound is given by: $c=\sqrt{gn}$ where $n$ is the local density. Now, in the presence of the imbalance potential, the right and left atom numbers are: $N_{1,2}=N(1\pm z_0)/2$, where the initial imbalance $z_0=(N_1-N_2)/N$ is normalized to the total atom number $N$, as in the main text. The associated speed of sounds are: $c_{1,2}=\sqrt{gn_0(1\pm z_0)}$ and assuming $z_0\ll1$ we obtain: $c_{\rm max}-c_{\rm min}=c_1-c_2\simeq\sqrt{gN/L}z_0$, corresponding to the main text formula.

\begin{figure}
    \centering
    \includegraphics[width=8.6cm]{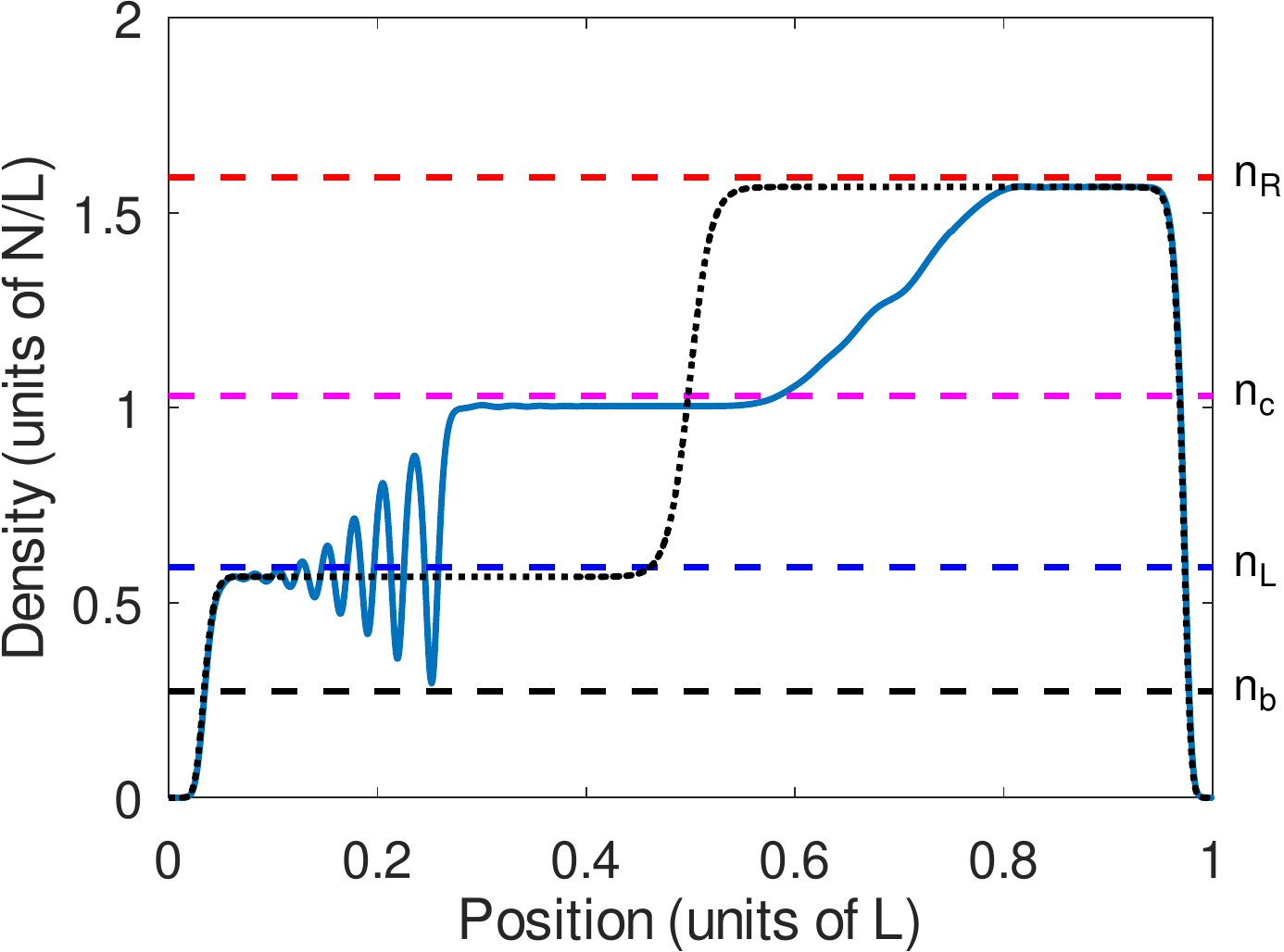}
    \caption{\label{fig:3levels} (color online) Density profile in the box at time $t=1.5\times10^{-3}\times mL^2/\hbar$ (solid blue line) for a initial dam-break problem (dotted black line), with $V_1=0$ and $V_{\rm offset}=gN$. The four horizontal dashed lines correspond to the analytical formula for $n_R$, $n_c$, $n_L$, $n_b$, respectively. The small discrepancy with the numerical solution may be attributed to finite size effects.}
\end{figure}

We now discuss the nucleation mechanism of individual solitons at the central barrier, for large initial imbalance. The initial state is very similar to the one of the dam break problem~\cite{El2016a,Xu2017} in dispersive shock waves. The time evolution involves the appearance of an intermediate plateau at a density: $n_c=(\sqrt{n_L}+\sqrt{n_R})^2/4$,
where $n_L < n_R$ are the initial left/right densities, see Fig.~\ref{fig:3levels}. Modulation theory then predicts that the density jump $n_c\to n_L$ is made of a soliton train and that the lowest density point of the oscillations is: $n_b=(\sqrt{n_c}-2\sqrt{n_L})^2$, see Eq. (3.55) of Ref.~\cite{El2016a}. Finally, assuming local density approximation, we can expect that additional individual solitons will be nucleated at the barrier when the local chemical potential vanishes, which occurs first when $V_1^s\simeq gn_b$, resulting in:
\begin{equation}
\label{eqn:V1s}
V_1^s=\mu\frac{\left(\sqrt{1+z_0}-3\sqrt{1-z_0}\right)^2}{4}.
\end{equation}

\section{\label{app:tmm}Two mode model}
We summarize here the derivation of a two mode model capturing the BJJ dynamics, following the approach of~\cite{Raghavan1999}. The idea is to look for a solution of Eq.~\eqref{dGPE} of the form:
\[
\psi(x,t)=\psi_1(t)\phi_1(x)+\psi_2(t)\phi_2(x),
\]
where the two modes $\phi_{1,2}(x)$ are build by combining the lowest symmetric $\phi_s(x)$ and anti-symmetric $\phi_a(x)$ energy states of Eq.~\eqref{GPE}:
\[
\phi_1(x)=\frac{\phi_s(x)+\phi_a(x)}{\sqrt{2}}~\textrm{and}~\phi_2(x)=\frac{\phi_s(x)-\phi_a(x)}{\sqrt{2}}.
\]
This definition ensures that $\phi_{1,2}(x)$ are normalized to unity and orthogonal. Without loss of generality we may assume that $\phi_{1,2}(x)$ are real valued functions.
Inserting this ansatz into Eq.~\eqref{dGPE} and projecting onto the modes result in the following equations for the mode amplitudes $\psi_{1,2}(t)$:
\begin{eqnarray*}
i\dot{\psi}_1&=&I_{11}\psi_1+I_{12}\psi_2+gN\int dx\,\phi_1|\psi|^2\psi,\\
i\dot{\psi}_2&=&I_{21}\psi_1+I_{22}\psi_2+gN\int dx\,\phi_2|\psi|^2\psi,
\end{eqnarray*}
where we defined the overlap integrals $I_{i,j}=\int dx\,\phi_i\left(-\frac{1}{2}\frac{\partial^2}{\partial x^2}+V(x)\right)\phi_j$. Here $V(x)$ is the final potential, with a central barrier of strength $V_1$.
To proceed we need to evaluate the non-linear term: we keep only the dominant term and the first correction~\footnote{In particular we neglect the anomalous terms involving the fields complex conjugates, as is usual in such expansions.}, resulting in:
\begin{eqnarray*}
i\dot{\psi}_1&=&I_{11}\psi_1+K\psi_2+U_1|\psi_1|^2\psi_1,\\
i\dot{\psi}_2&=&K\psi_1+I_{22}\psi_2+U_2|\psi_2|^2\psi_2,
\end{eqnarray*}
where we defined $U_{1,2}=gN\int dx\,|\phi_{1,2}|^4$ and introduced the coupling:
\[
K=I_{12}+\frac{gN}{4}\int dx\,(\phi_s^4-\phi_a^4).
\]
It is then straightforward to show that:
\[
K=\frac{E[\phi_s]-E[\phi_a]}{2},
\]
where
\begin{equation}
E[\phi]=\int dx\,\phi\left[-\frac{1}{2}\frac{\partial^2}{\partial x^2}+V(x)+\frac{gN}{2}|\phi|^2\right]\phi,
\label{eqn:energy}
\end{equation}
is the mean field energy of the state with $N$ particles in mode $\phi$.

We then change variables for $\psi_{1,2}=\sqrt{N_{1,2}}e^{-i\theta_{1,2}}$ and obtain:
\begin{eqnarray*}
\dot{N}_1&=&2K\sqrt{N_1N_2}\sin{(\theta_1-\theta_2)},\\
\dot{\theta}_1&=&I_{11}+K\sqrt{\frac{N_2}{N_1}}\cos{(\theta_1-\theta_2)}+U_1N_1,\\
\dot{N}_2&=&2K\sqrt{N_1N_2}\sin{(\theta_2-\theta_1)},\\
\dot{\theta}_2&=&K\sqrt{\frac{N_1}{N_2}}\cos{(\theta_2-\theta_1)}+I_{22}+U_2N_2.
\end{eqnarray*}
We then introduce the population imbalance: $z=N_1-N_2$ (with the constrain $N_1+N_2=1$, due to the choice of mode function normalization) and the phase difference $\theta=\theta_1-\theta_2$, resulting in:
\begin{eqnarray*}
\dot{z}&=&2K\sqrt{1-z^2}\sin{\theta},\\
\dot{\theta}&=&\left[U-2K\frac{\cos{\theta}}{\sqrt{1-z^2}}\right]z+I_{11}-I_{22}+\frac{U_1-U_2}{2},
\end{eqnarray*}
where $U=(U_1+U_2)/2$.
The constant term in the second equation is very small such that we recover the usual Bose-Josephson equations (with our definitions $K<0$), as shown in Eq.~\ref{eqn:tmmstd}.

To connect the GPE simulations and the two mode model, we minimize the energy~\eqref{eqn:energy} for a given central barrier $V_1$ with a parity constrain~\footnote{This constrain is naturally implemented with a spectral scheme using the discrete cosine (sine) transform to find even (odd) states.} to find the two states $\phi_s$ and $\phi_a$ and their energy, thus obtaining the value of $K$. We then combine them to build states $\phi_1$ and $\phi_2$ and compute the interaction energy $U$. Combining $K$ and $U$ we obtain the parameter $\Lambda=U/(2|K|)$ controlling the dynamics as a function of $V_1$.
Finally, using the fact that the transition from oscillations to self-trapping occurs at a critical $\Lambda$~\cite{Raghavan1999}:
\begin{equation}
\Lambda_c=\frac{1+\sqrt{1-z_0^2}\cos{\theta_0}}{z_0^2/2},
\label{eqn:critical}
\end{equation}
where $z_0$ and $\theta_0$ are the initial imbalance and phase difference between the reservoirs. As each value of $\Lambda$ correspond to a unique value of $V_1$, Eq.~\eqref{eqn:critical} allows to plot the two-mode prediction for the JO to ST boundary in Fig.~\ref{fig:PSD}.

\section{\label{app:tau}Additional data for the barrier quench}
\begin{figure}[tb]
\centering
\includegraphics[width=8.6cm]{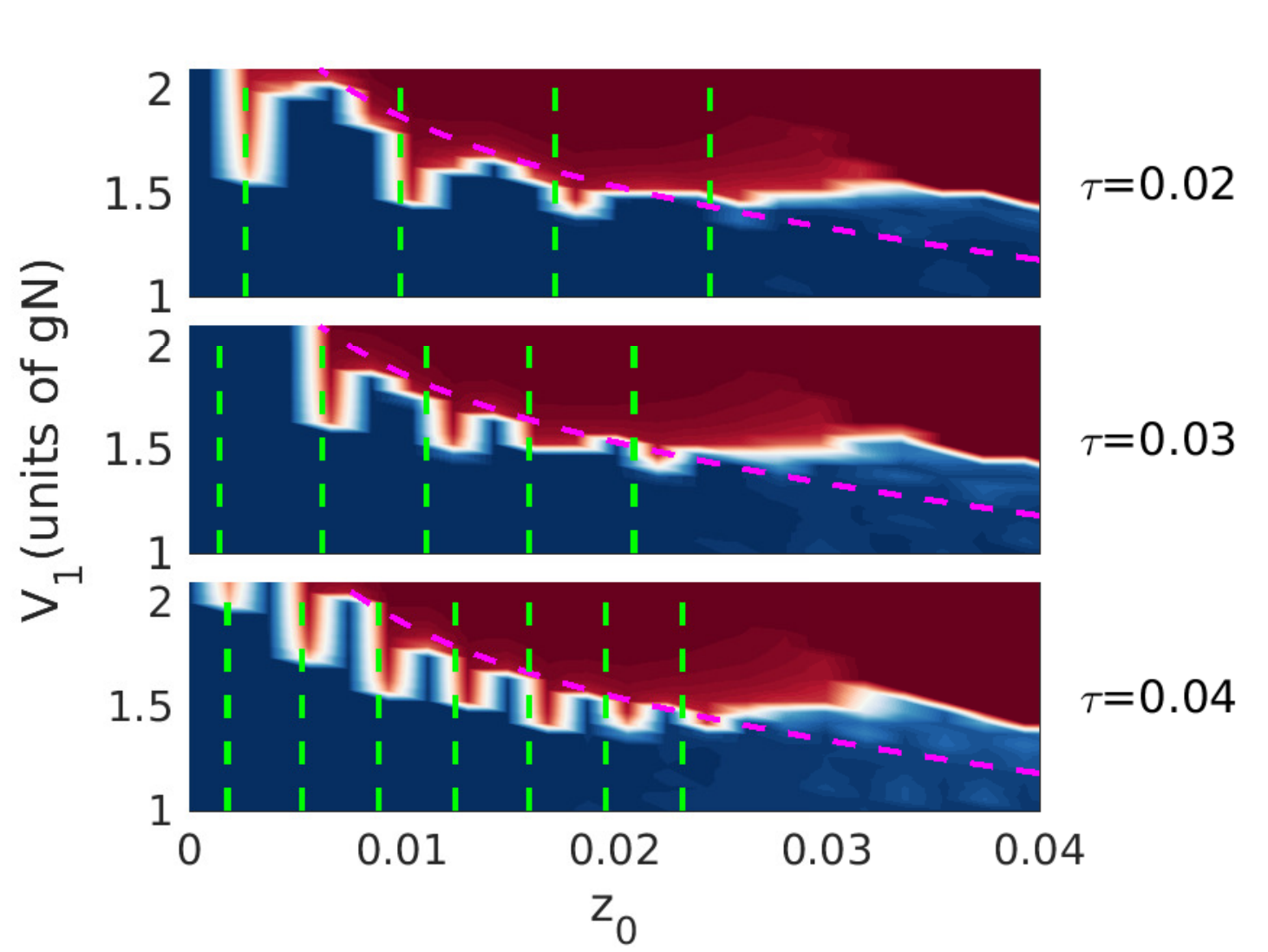} 
\caption{
\label{fig:extra_tau}
Normalized power spectrum density at zero frequency as a function of central barrier height $V_1$ and initial imbalance $z_0$ for a quench of $\tau =\{0.02,0.03,0.04\}$ (top, middle, bottom). The magenta dashed line indicates the prediction of the two-mode model for the self-trapping transition. The dashed vertical green lines correspond to the predicted resonances.
}
\end{figure}
Figure~\ref{fig:extra_tau} shows how the quench time $\tau$ affects the self trapping resonance described in the main text. In particular it confirms the interpretation based on the two-mode model equations and show the predictive powers of this simple model. We note that the number of resonances increases with $\tau$ but are also less pronounced. For $\tau>0.05$ they are indistinguishable from the two-mode model boundary, shown by the dashed magenta curve in Fig.~\ref{fig:extra_tau}.

\begin{figure*}[tb]
\centering
\includegraphics[width=17.8cm]{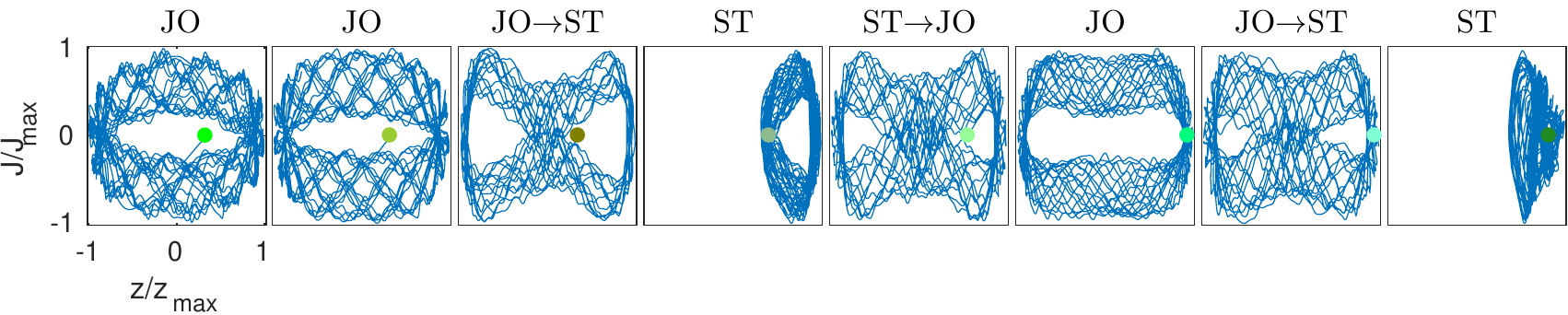}
\caption{\label{fig:extraTraj}
Trajectories in the $J(t)$ versus $z(t)$ plane, normalized to the maximal current $J_{\rm max}$ and maximal imbalance $z_{\rm max}$, corresponding to the examples shown in Fig.~\ref{fig:tau:0.01}. From left to right: increasing initial imbalance, going from the JO regime to ST, then back to JO and finally entering the ST regime again. In all phase portrait the green disk indicate the initial condition ($z=z_0$ and $J=0$).
}
\end{figure*}


\begin{thebibliography}{65}%
\makeatletter
\providecommand \@ifxundefined [1]{%
 \@ifx{#1\undefined}
}%
\providecommand \@ifnum [1]{%
 \ifnum #1\expandafter \@firstoftwo
 \else \expandafter \@secondoftwo
 \fi
}%
\providecommand \@ifx [1]{%
 \ifx #1\expandafter \@firstoftwo
 \else \expandafter \@secondoftwo
 \fi
}%
\providecommand \natexlab [1]{#1}%
\providecommand \enquote  [1]{``#1''}%
\providecommand \bibnamefont  [1]{#1}%
\providecommand \bibfnamefont [1]{#1}%
\providecommand \citenamefont [1]{#1}%
\providecommand \href@noop [0]{\@secondoftwo}%
\providecommand \href [0]{\begingroup \@sanitize@url \@href}%
\providecommand \@href[1]{\@@startlink{#1}\@@href}%
\providecommand \@@href[1]{\endgroup#1\@@endlink}%
\providecommand \@sanitize@url [0]{\catcode `\\12\catcode `\$12\catcode
  `\&12\catcode `\#12\catcode `\^12\catcode `\_12\catcode `\%12\relax}%
\providecommand \@@startlink[1]{}%
\providecommand \@@endlink[0]{}%
\providecommand \url  [0]{\begingroup\@sanitize@url \@url }%
\providecommand \@url [1]{\endgroup\@href {#1}{\urlprefix }}%
\providecommand \urlprefix  [0]{URL }%
\providecommand \Eprint [0]{\href }%
\providecommand \doibase [0]{https://doi.org/}%
\providecommand \selectlanguage [0]{\@gobble}%
\providecommand \bibinfo  [0]{\@secondoftwo}%
\providecommand \bibfield  [0]{\@secondoftwo}%
\providecommand \translation [1]{[#1]}%
\providecommand \BibitemOpen [0]{}%
\providecommand \bibitemStop [0]{}%
\providecommand \bibitemNoStop [0]{.\EOS\space}%
\providecommand \EOS [0]{\spacefactor3000\relax}%
\providecommand \BibitemShut  [1]{\csname bibitem#1\endcsname}%
\let\auto@bib@innerbib\@empty
\bibitem [{\citenamefont {Josephson}(1962)}]{Josephson1962}%
  \BibitemOpen
  \bibfield  {author} {\bibinfo {author} {\bibfnamefont {B.~D.}\ \bibnamefont
  {Josephson}},\ }\bibfield  {title} {\bibinfo {title} {Possible new effects in
  superconductive tunnelling},\ }\href
  {https://doi.org/https://doi.org/10.1016/0031-9163(62)91369-0} {\bibfield
  {journal} {\bibinfo  {journal} {Physics Letters}\ }\textbf {\bibinfo {volume}
  {1}},\ \bibinfo {pages} {251 } (\bibinfo {year} {1962})}\BibitemShut
  {NoStop}%
\bibitem [{\citenamefont {Anderson}\ and\ \citenamefont
  {Rowell}(1963)}]{Anderson1963}%
  \BibitemOpen
  \bibfield  {author} {\bibinfo {author} {\bibfnamefont {P.~W.}\ \bibnamefont
  {Anderson}}\ and\ \bibinfo {author} {\bibfnamefont {J.~M.}\ \bibnamefont
  {Rowell}},\ }\bibfield  {title} {\bibinfo {title} {Probable observation of
  the josephson superconducting tunneling effect},\ }\href
  {https://doi.org/10.1103/PhysRevLett.10.230} {\bibfield  {journal} {\bibinfo
  {journal} {Phys. Rev. Lett.}\ }\textbf {\bibinfo {volume} {10}},\ \bibinfo
  {pages} {230} (\bibinfo {year} {1963})}\BibitemShut {NoStop}%
\bibitem [{\citenamefont {Likharev}(1979)}]{Likharev1979}%
  \BibitemOpen
  \bibfield  {author} {\bibinfo {author} {\bibfnamefont {K.~K.}\ \bibnamefont
  {Likharev}},\ }\bibfield  {title} {\bibinfo {title} {Superconducting weak
  links},\ }\href {https://doi.org/10.1103/RevModPhys.51.101} {\bibfield
  {journal} {\bibinfo  {journal} {Rev. Mod. Phys.}\ }\textbf {\bibinfo {volume}
  {51}},\ \bibinfo {pages} {101} (\bibinfo {year} {1979})}\BibitemShut
  {NoStop}%
\bibitem [{\citenamefont {Barone}\ and\ \citenamefont
  {Paterno}(1982)}]{Barone1982}%
  \BibitemOpen
  \bibfield  {author} {\bibinfo {author} {\bibfnamefont {A.}~\bibnamefont
  {Barone}}\ and\ \bibinfo {author} {\bibfnamefont {G.}~\bibnamefont
  {Paterno}},\ }\href@noop {} {\emph {\bibinfo {title} {Physics and
  Applications of the Josephson Effect}}}\ (\bibinfo  {publisher} {Wiley},\
  \bibinfo {address} {New York},\ \bibinfo {year} {1982})\BibitemShut {NoStop}%
\bibitem [{\citenamefont {Smerzi}\ \emph {et~al.}(1997)\citenamefont {Smerzi},
  \citenamefont {Fantoni}, \citenamefont {Giovanazzi},\ and\ \citenamefont
  {Shenoy}}]{Smerzi1997}%
  \BibitemOpen
  \bibfield  {author} {\bibinfo {author} {\bibfnamefont {A.}~\bibnamefont
  {Smerzi}}, \bibinfo {author} {\bibfnamefont {S.}~\bibnamefont {Fantoni}},
  \bibinfo {author} {\bibfnamefont {S.}~\bibnamefont {Giovanazzi}},\ and\
  \bibinfo {author} {\bibfnamefont {S.~R.}\ \bibnamefont {Shenoy}},\ }\bibfield
   {title} {\bibinfo {title} {Quantum coherent atomic tunneling between two
  trapped bose-einstein condensates},\ }\href
  {https://doi.org/10.1103/PhysRevLett.79.4950} {\bibfield  {journal} {\bibinfo
   {journal} {Phys. Rev. Lett.}\ }\textbf {\bibinfo {volume} {79}},\ \bibinfo
  {pages} {4950} (\bibinfo {year} {1997})}\BibitemShut {NoStop}%
\bibitem [{\citenamefont {Milburn}\ \emph {et~al.}(1997)\citenamefont
  {Milburn}, \citenamefont {Corney}, \citenamefont {Wright},\ and\
  \citenamefont {Walls}}]{Milburn1997}%
  \BibitemOpen
  \bibfield  {author} {\bibinfo {author} {\bibfnamefont {G.~J.}\ \bibnamefont
  {Milburn}}, \bibinfo {author} {\bibfnamefont {J.}~\bibnamefont {Corney}},
  \bibinfo {author} {\bibfnamefont {E.~M.}\ \bibnamefont {Wright}},\ and\
  \bibinfo {author} {\bibfnamefont {D.~F.}\ \bibnamefont {Walls}},\ }\bibfield
  {title} {\bibinfo {title} {Quantum dynamics of an atomic bose-einstein
  condensate in a double-well potential},\ }\href
  {https://doi.org/10.1103/PhysRevA.55.4318} {\bibfield  {journal} {\bibinfo
  {journal} {Phys. Rev. A}\ }\textbf {\bibinfo {volume} {55}},\ \bibinfo
  {pages} {4318} (\bibinfo {year} {1997})}\BibitemShut {NoStop}%
\bibitem [{\citenamefont {Zapata}\ \emph {et~al.}(1998)\citenamefont {Zapata},
  \citenamefont {Sols},\ and\ \citenamefont {Leggett}}]{Zapata1998}%
  \BibitemOpen
  \bibfield  {author} {\bibinfo {author} {\bibfnamefont {I.}~\bibnamefont
  {Zapata}}, \bibinfo {author} {\bibfnamefont {F.}~\bibnamefont {Sols}},\ and\
  \bibinfo {author} {\bibfnamefont {A.~J.}\ \bibnamefont {Leggett}},\
  }\bibfield  {title} {\bibinfo {title} {Josephson effect between trapped
  bose-einstein condensates},\ }\href {https://doi.org/10.1103/PhysRevA.57.R28}
  {\bibfield  {journal} {\bibinfo  {journal} {Phys. Rev. A}\ }\textbf {\bibinfo
  {volume} {57}},\ \bibinfo {pages} {R28} (\bibinfo {year} {1998})}\BibitemShut
  {NoStop}%
\bibitem [{\citenamefont {Raghavan}\ \emph
  {et~al.}(1999{\natexlab{a}})\citenamefont {Raghavan}, \citenamefont {Smerzi},
  \citenamefont {Fantoni},\ and\ \citenamefont {Shenoy}}]{Raghavan1999}%
  \BibitemOpen
  \bibfield  {author} {\bibinfo {author} {\bibfnamefont {S.}~\bibnamefont
  {Raghavan}}, \bibinfo {author} {\bibfnamefont {A.}~\bibnamefont {Smerzi}},
  \bibinfo {author} {\bibfnamefont {S.}~\bibnamefont {Fantoni}},\ and\ \bibinfo
  {author} {\bibfnamefont {S.~R.}\ \bibnamefont {Shenoy}},\ }\bibfield  {title}
  {\bibinfo {title} {Coherent oscillations between two weakly coupled
  bose-einstein condensates: Josephson effects, $\ensuremath{\pi}$
  oscillations, and macroscopic quantum self-trapping},\ }\href
  {https://doi.org/10.1103/PhysRevA.59.620} {\bibfield  {journal} {\bibinfo
  {journal} {Phys. Rev. A}\ }\textbf {\bibinfo {volume} {59}},\ \bibinfo
  {pages} {620} (\bibinfo {year} {1999}{\natexlab{a}})}\BibitemShut {NoStop}%
\bibitem [{\citenamefont {Raghavan}\ \emph
  {et~al.}(1999{\natexlab{b}})\citenamefont {Raghavan}, \citenamefont
  {Smerzi},\ and\ \citenamefont {Kenkre}}]{Raghavan1999b}%
  \BibitemOpen
  \bibfield  {author} {\bibinfo {author} {\bibfnamefont {S.}~\bibnamefont
  {Raghavan}}, \bibinfo {author} {\bibfnamefont {A.}~\bibnamefont {Smerzi}},\
  and\ \bibinfo {author} {\bibfnamefont {V.~M.}\ \bibnamefont {Kenkre}},\
  }\bibfield  {title} {\bibinfo {title} {Transitions in coherent oscillations
  between two trapped bose-einstein condensates},\ }\href
  {https://doi.org/10.1103/PhysRevA.60.R1787} {\bibfield  {journal} {\bibinfo
  {journal} {Phys. Rev. A}\ }\textbf {\bibinfo {volume} {60}},\ \bibinfo
  {pages} {R1787} (\bibinfo {year} {1999}{\natexlab{b}})}\BibitemShut {NoStop}%
\bibitem [{\citenamefont {Cataliotti}\ \emph {et~al.}(2001)\citenamefont
  {Cataliotti}, \citenamefont {Burger}, \citenamefont {Fort}, \citenamefont
  {Maddaloni}, \citenamefont {Minardi}, \citenamefont {Trombettoni},
  \citenamefont {Smerzi},\ and\ \citenamefont {Inguscio}}]{Cataliotti2001}%
  \BibitemOpen
  \bibfield  {author} {\bibinfo {author} {\bibfnamefont {F.~S.}\ \bibnamefont
  {Cataliotti}}, \bibinfo {author} {\bibfnamefont {S.}~\bibnamefont {Burger}},
  \bibinfo {author} {\bibfnamefont {C.}~\bibnamefont {Fort}}, \bibinfo {author}
  {\bibfnamefont {P.}~\bibnamefont {Maddaloni}}, \bibinfo {author}
  {\bibfnamefont {F.}~\bibnamefont {Minardi}}, \bibinfo {author} {\bibfnamefont
  {A.}~\bibnamefont {Trombettoni}}, \bibinfo {author} {\bibfnamefont
  {A.}~\bibnamefont {Smerzi}},\ and\ \bibinfo {author} {\bibfnamefont
  {M.}~\bibnamefont {Inguscio}},\ }\bibfield  {title} {\bibinfo {title}
  {Josephson junction arrays with bose-einstein condensates},\ }\href
  {https://doi.org/10.1126/science.1062612} {\bibfield  {journal} {\bibinfo
  {journal} {Science}\ }\textbf {\bibinfo {volume} {293}},\ \bibinfo {pages}
  {843} (\bibinfo {year} {2001})}\BibitemShut {NoStop}%
\bibitem [{\citenamefont {Albiez}\ \emph {et~al.}(2005)\citenamefont {Albiez},
  \citenamefont {Gati}, \citenamefont {F\"olling}, \citenamefont {Hunsmann},
  \citenamefont {Cristiani},\ and\ \citenamefont {Oberthaler}}]{Albiez2005}%
  \BibitemOpen
  \bibfield  {author} {\bibinfo {author} {\bibfnamefont {M.}~\bibnamefont
  {Albiez}}, \bibinfo {author} {\bibfnamefont {R.}~\bibnamefont {Gati}},
  \bibinfo {author} {\bibfnamefont {J.}~\bibnamefont {F\"olling}}, \bibinfo
  {author} {\bibfnamefont {S.}~\bibnamefont {Hunsmann}}, \bibinfo {author}
  {\bibfnamefont {M.}~\bibnamefont {Cristiani}},\ and\ \bibinfo {author}
  {\bibfnamefont {M.~K.}\ \bibnamefont {Oberthaler}},\ }\bibfield  {title}
  {\bibinfo {title} {Direct observation of tunneling and nonlinear
  self-trapping in a single bosonic josephson junction},\ }\href
  {https://doi.org/10.1103/PhysRevLett.95.010402} {\bibfield  {journal}
  {\bibinfo  {journal} {Phys. Rev. Lett.}\ }\textbf {\bibinfo {volume} {95}},\
  \bibinfo {pages} {010402} (\bibinfo {year} {2005})}\BibitemShut {NoStop}%
\bibitem [{\citenamefont {Levy}\ \emph {et~al.}(2007)\citenamefont {Levy},
  \citenamefont {Lahoud}, \citenamefont {Shomroni},\ and\ \citenamefont
  {Steinhauer}}]{Levy2007}%
  \BibitemOpen
  \bibfield  {author} {\bibinfo {author} {\bibfnamefont {S.}~\bibnamefont
  {Levy}}, \bibinfo {author} {\bibfnamefont {E.}~\bibnamefont {Lahoud}},
  \bibinfo {author} {\bibfnamefont {I.}~\bibnamefont {Shomroni}},\ and\
  \bibinfo {author} {\bibfnamefont {J.}~\bibnamefont {Steinhauer}},\ }\bibfield
   {title} {\bibinfo {title} {The a.c. and d.c. josephson effects in a
  bose–einstein condensate},\ }\href {https://doi.org/10.1038/nature06186}
  {\bibfield  {journal} {\bibinfo  {journal} {Nature}\ }\textbf {\bibinfo
  {volume} {449}},\ \bibinfo {pages} {579} (\bibinfo {year}
  {2007})}\BibitemShut {NoStop}%
\bibitem [{\citenamefont {Meier}\ and\ \citenamefont
  {Zwerger}(2001)}]{Meier2001}%
  \BibitemOpen
  \bibfield  {author} {\bibinfo {author} {\bibfnamefont {F.}~\bibnamefont
  {Meier}}\ and\ \bibinfo {author} {\bibfnamefont {W.}~\bibnamefont
  {Zwerger}},\ }\bibfield  {title} {\bibinfo {title} {Josephson tunneling
  between weakly interacting bose-einstein condensates},\ }\href
  {https://doi.org/10.1103/PhysRevA.64.033610} {\bibfield  {journal} {\bibinfo
  {journal} {Phys. Rev. A}\ }\textbf {\bibinfo {volume} {64}},\ \bibinfo
  {pages} {033610} (\bibinfo {year} {2001})}\BibitemShut {NoStop}%
\bibitem [{\citenamefont {LeBlanc}\ \emph {et~al.}(2011)\citenamefont
  {LeBlanc}, \citenamefont {Bardon}, \citenamefont {McKeever}, \citenamefont
  {Extavour}, \citenamefont {Jervis}, \citenamefont {Thywissen}, \citenamefont
  {Piazza},\ and\ \citenamefont {Smerzi}}]{LeBlanc2011}%
  \BibitemOpen
  \bibfield  {author} {\bibinfo {author} {\bibfnamefont {L.~J.}\ \bibnamefont
  {LeBlanc}}, \bibinfo {author} {\bibfnamefont {A.~B.}\ \bibnamefont {Bardon}},
  \bibinfo {author} {\bibfnamefont {J.}~\bibnamefont {McKeever}}, \bibinfo
  {author} {\bibfnamefont {M.~H.~T.}\ \bibnamefont {Extavour}}, \bibinfo
  {author} {\bibfnamefont {D.}~\bibnamefont {Jervis}}, \bibinfo {author}
  {\bibfnamefont {J.~H.}\ \bibnamefont {Thywissen}}, \bibinfo {author}
  {\bibfnamefont {F.}~\bibnamefont {Piazza}},\ and\ \bibinfo {author}
  {\bibfnamefont {A.}~\bibnamefont {Smerzi}},\ }\bibfield  {title} {\bibinfo
  {title} {Dynamics of a tunable superfluid junction},\ }\href
  {https://doi.org/10.1103/PhysRevLett.106.025302} {\bibfield  {journal}
  {\bibinfo  {journal} {Phys. Rev. Lett.}\ }\textbf {\bibinfo {volume} {106}},\
  \bibinfo {pages} {025302} (\bibinfo {year} {2011})}\BibitemShut {NoStop}%
\bibitem [{\citenamefont {Pigneur}\ \emph {et~al.}(2018)\citenamefont
  {Pigneur}, \citenamefont {Berrada}, \citenamefont {Bonneau}, \citenamefont
  {Schumm}, \citenamefont {Demler},\ and\ \citenamefont
  {Schmiedmayer}}]{Pigneur2018}%
  \BibitemOpen
  \bibfield  {author} {\bibinfo {author} {\bibfnamefont {M.}~\bibnamefont
  {Pigneur}}, \bibinfo {author} {\bibfnamefont {T.}~\bibnamefont {Berrada}},
  \bibinfo {author} {\bibfnamefont {M.}~\bibnamefont {Bonneau}}, \bibinfo
  {author} {\bibfnamefont {T.}~\bibnamefont {Schumm}}, \bibinfo {author}
  {\bibfnamefont {E.}~\bibnamefont {Demler}},\ and\ \bibinfo {author}
  {\bibfnamefont {J.}~\bibnamefont {Schmiedmayer}},\ }\bibfield  {title}
  {\bibinfo {title} {Relaxation to a phase-locked equilibrium state in a
  one-dimensional bosonic josephson junction},\ }\href
  {https://doi.org/10.1103/PhysRevLett.120.173601} {\bibfield  {journal}
  {\bibinfo  {journal} {Phys. Rev. Lett.}\ }\textbf {\bibinfo {volume} {120}},\
  \bibinfo {pages} {173601} (\bibinfo {year} {2018})}\BibitemShut {NoStop}%
\bibitem [{\citenamefont {Saha}\ \emph {et~al.}(2020)\citenamefont {Saha},
  \citenamefont {Ray},\ and\ \citenamefont {Deb}}]{Saha2020}%
  \BibitemOpen
  \bibfield  {author} {\bibinfo {author} {\bibfnamefont {A.~K.}\ \bibnamefont
  {Saha}}, \bibinfo {author} {\bibfnamefont {D.~S.}\ \bibnamefont {Ray}},\ and\
  \bibinfo {author} {\bibfnamefont {B.}~\bibnamefont {Deb}},\ }\bibfield
  {title} {\bibinfo {title} {{Parametric oscillations in a dissipative bosonic
  Josephson junction}},\ }\href {https://doi.org/10.1088/1361-6455/ab82e3}
  {\bibfield  {journal} {\bibinfo  {journal} {Journal of Physics B: Atomic,
  Molecular and Optical Physics}\ }\textbf {\bibinfo {volume} {53}},\ \bibinfo
  {pages} {135301} (\bibinfo {year} {2020})}\BibitemShut {NoStop}%
\bibitem [{\citenamefont {Smerzi}\ \emph {et~al.}(2003)\citenamefont {Smerzi},
  \citenamefont {Trombettoni}, \citenamefont {Lopez-Arias}, \citenamefont
  {Fort}, \citenamefont {Maddaloni}, \citenamefont {Minardi},\ and\
  \citenamefont {Inguscio}}]{Smerzi2003}%
  \BibitemOpen
  \bibfield  {author} {\bibinfo {author} {\bibfnamefont {A.}~\bibnamefont
  {Smerzi}}, \bibinfo {author} {\bibfnamefont {A.}~\bibnamefont {Trombettoni}},
  \bibinfo {author} {\bibfnamefont {T.}~\bibnamefont {Lopez-Arias}}, \bibinfo
  {author} {\bibfnamefont {C.}~\bibnamefont {Fort}}, \bibinfo {author}
  {\bibfnamefont {P.}~\bibnamefont {Maddaloni}}, \bibinfo {author}
  {\bibfnamefont {F.}~\bibnamefont {Minardi}},\ and\ \bibinfo {author}
  {\bibfnamefont {M.}~\bibnamefont {Inguscio}},\ }\bibfield  {title} {\bibinfo
  {title} {Macroscopic oscillations between two weakly coupled bose-einstein
  condensates},\ }\href {https://doi.org/10.1140/epjb/e2003-00055-1} {\bibfield
   {journal} {\bibinfo  {journal} {The European Physical Journal B - Condensed
  Matter and Complex Systems}\ }\textbf {\bibinfo {volume} {31}},\ \bibinfo
  {pages} {457} (\bibinfo {year} {2003})}\BibitemShut {NoStop}%
\bibitem [{\citenamefont {Spagnolli}\ \emph {et~al.}(2017)\citenamefont
  {Spagnolli}, \citenamefont {Semeghini}, \citenamefont {Masi}, \citenamefont
  {Ferioli}, \citenamefont {Trenkwalder}, \citenamefont {Coop}, \citenamefont
  {Landini}, \citenamefont {Pezz\`e}, \citenamefont {Modugno}, \citenamefont
  {Inguscio}, \citenamefont {Smerzi},\ and\ \citenamefont
  {Fattori}}]{Spagnolli2017}%
  \BibitemOpen
  \bibfield  {author} {\bibinfo {author} {\bibfnamefont {G.}~\bibnamefont
  {Spagnolli}}, \bibinfo {author} {\bibfnamefont {G.}~\bibnamefont
  {Semeghini}}, \bibinfo {author} {\bibfnamefont {L.}~\bibnamefont {Masi}},
  \bibinfo {author} {\bibfnamefont {G.}~\bibnamefont {Ferioli}}, \bibinfo
  {author} {\bibfnamefont {A.}~\bibnamefont {Trenkwalder}}, \bibinfo {author}
  {\bibfnamefont {S.}~\bibnamefont {Coop}}, \bibinfo {author} {\bibfnamefont
  {M.}~\bibnamefont {Landini}}, \bibinfo {author} {\bibfnamefont
  {L.}~\bibnamefont {Pezz\`e}}, \bibinfo {author} {\bibfnamefont
  {G.}~\bibnamefont {Modugno}}, \bibinfo {author} {\bibfnamefont
  {M.}~\bibnamefont {Inguscio}}, \bibinfo {author} {\bibfnamefont
  {A.}~\bibnamefont {Smerzi}},\ and\ \bibinfo {author} {\bibfnamefont
  {M.}~\bibnamefont {Fattori}},\ }\bibfield  {title} {\bibinfo {title}
  {Crossing over from attractive to repulsive interactions in a tunneling
  bosonic josephson junction},\ }\href
  {https://doi.org/10.1103/PhysRevLett.118.230403} {\bibfield  {journal}
  {\bibinfo  {journal} {Phys. Rev. Lett.}\ }\textbf {\bibinfo {volume} {118}},\
  \bibinfo {pages} {230403} (\bibinfo {year} {2017})}\BibitemShut {NoStop}%
\bibitem [{\citenamefont {Mart\'{\i}nez-Garaot}\ \emph
  {et~al.}(2018)\citenamefont {Mart\'{\i}nez-Garaot}, \citenamefont {Pettini},\
  and\ \citenamefont {Modugno}}]{MartinezGaraot2018}%
  \BibitemOpen
  \bibfield  {author} {\bibinfo {author} {\bibfnamefont {S.}~\bibnamefont
  {Mart\'{\i}nez-Garaot}}, \bibinfo {author} {\bibfnamefont {G.}~\bibnamefont
  {Pettini}},\ and\ \bibinfo {author} {\bibfnamefont {M.}~\bibnamefont
  {Modugno}},\ }\bibfield  {title} {\bibinfo {title} {Nonlinear mixing of
  bogoliubov modes in a bosonic josephson junction},\ }\href
  {https://doi.org/10.1103/PhysRevA.98.043624} {\bibfield  {journal} {\bibinfo
  {journal} {Phys. Rev. A}\ }\textbf {\bibinfo {volume} {98}},\ \bibinfo
  {pages} {043624} (\bibinfo {year} {2018})}\BibitemShut {NoStop}%
\bibitem [{\citenamefont {Ramanathan}\ \emph {et~al.}(2011)\citenamefont
  {Ramanathan}, \citenamefont {Wright}, \citenamefont {Muniz}, \citenamefont
  {Zelan}, \citenamefont {Hill}, \citenamefont {Lobb}, \citenamefont
  {Helmerson}, \citenamefont {Phillips},\ and\ \citenamefont
  {Campbell}}]{Ramanathan2011}%
  \BibitemOpen
  \bibfield  {author} {\bibinfo {author} {\bibfnamefont {A.}~\bibnamefont
  {Ramanathan}}, \bibinfo {author} {\bibfnamefont {K.~C.}\ \bibnamefont
  {Wright}}, \bibinfo {author} {\bibfnamefont {S.~R.}\ \bibnamefont {Muniz}},
  \bibinfo {author} {\bibfnamefont {M.}~\bibnamefont {Zelan}}, \bibinfo
  {author} {\bibfnamefont {W.~T.}\ \bibnamefont {Hill}}, \bibinfo {author}
  {\bibfnamefont {C.~J.}\ \bibnamefont {Lobb}}, \bibinfo {author}
  {\bibfnamefont {K.}~\bibnamefont {Helmerson}}, \bibinfo {author}
  {\bibfnamefont {W.~D.}\ \bibnamefont {Phillips}},\ and\ \bibinfo {author}
  {\bibfnamefont {G.~K.}\ \bibnamefont {Campbell}},\ }\bibfield  {title}
  {\bibinfo {title} {Superflow in a toroidal bose-einstein condensate: An atom
  circuit with a tunable weak link},\ }\href
  {https://doi.org/10.1103/PhysRevLett.106.130401} {\bibfield  {journal}
  {\bibinfo  {journal} {Phys. Rev. Lett.}\ }\textbf {\bibinfo {volume} {106}},\
  \bibinfo {pages} {130401} (\bibinfo {year} {2011})}\BibitemShut {NoStop}%
\bibitem [{\citenamefont {Ryu}\ \emph {et~al.}(2013)\citenamefont {Ryu},
  \citenamefont {Blackburn}, \citenamefont {Blinova},\ and\ \citenamefont
  {Boshier}}]{Ryu2013}%
  \BibitemOpen
  \bibfield  {author} {\bibinfo {author} {\bibfnamefont {C.}~\bibnamefont
  {Ryu}}, \bibinfo {author} {\bibfnamefont {P.~W.}\ \bibnamefont {Blackburn}},
  \bibinfo {author} {\bibfnamefont {A.~A.}\ \bibnamefont {Blinova}},\ and\
  \bibinfo {author} {\bibfnamefont {M.~G.}\ \bibnamefont {Boshier}},\
  }\bibfield  {title} {\bibinfo {title} {Experimental realization of josephson
  junctions for an atom squid},\ }\href
  {https://doi.org/10.1103/PhysRevLett.111.205301} {\bibfield  {journal}
  {\bibinfo  {journal} {Phys. Rev. Lett.}\ }\textbf {\bibinfo {volume} {111}},\
  \bibinfo {pages} {205301} (\bibinfo {year} {2013})}\BibitemShut {NoStop}%
\bibitem [{\citenamefont {Polo}\ \emph {et~al.}(2019)\citenamefont {Polo},
  \citenamefont {Dubessy}, \citenamefont {Pedri}, \citenamefont {Perrin},\ and\
  \citenamefont {Minguzzi}}]{Polo2019}%
  \BibitemOpen
  \bibfield  {author} {\bibinfo {author} {\bibfnamefont {J.}~\bibnamefont
  {Polo}}, \bibinfo {author} {\bibfnamefont {R.}~\bibnamefont {Dubessy}},
  \bibinfo {author} {\bibfnamefont {P.}~\bibnamefont {Pedri}}, \bibinfo
  {author} {\bibfnamefont {H.}~\bibnamefont {Perrin}},\ and\ \bibinfo {author}
  {\bibfnamefont {A.}~\bibnamefont {Minguzzi}},\ }\bibfield  {title} {\bibinfo
  {title} {Oscillations and decay of superfluid currents in a one-dimensional
  bose gas on a ring},\ }\href {https://doi.org/10.1103/PhysRevLett.123.195301}
  {\bibfield  {journal} {\bibinfo  {journal} {Phys. Rev. Lett.}\ }\textbf
  {\bibinfo {volume} {123}},\ \bibinfo {pages} {195301} (\bibinfo {year}
  {2019})}\BibitemShut {NoStop}%
\bibitem [{\citenamefont {Eckel}\ \emph {et~al.}(2014)\citenamefont {Eckel},
  \citenamefont {Jendrzejewski}, \citenamefont {Kumar}, \citenamefont {Lobb},\
  and\ \citenamefont {Campbell}}]{Eckel2014}%
  \BibitemOpen
  \bibfield  {author} {\bibinfo {author} {\bibfnamefont {S.}~\bibnamefont
  {Eckel}}, \bibinfo {author} {\bibfnamefont {F.}~\bibnamefont
  {Jendrzejewski}}, \bibinfo {author} {\bibfnamefont {A.}~\bibnamefont
  {Kumar}}, \bibinfo {author} {\bibfnamefont {C.~J.}\ \bibnamefont {Lobb}},\
  and\ \bibinfo {author} {\bibfnamefont {G.~K.}\ \bibnamefont {Campbell}},\
  }\bibfield  {title} {\bibinfo {title} {Interferometric measurement of the
  current-phase relationship of a superfluid weak link},\ }\href
  {https://doi.org/10.1103/PhysRevX.4.031052} {\bibfield  {journal} {\bibinfo
  {journal} {Phys. Rev. X}\ }\textbf {\bibinfo {volume} {4}},\ \bibinfo {pages}
  {031052} (\bibinfo {year} {2014})}\BibitemShut {NoStop}%
\bibitem [{\citenamefont {Jendrzejewski}\ \emph {et~al.}(2014)\citenamefont
  {Jendrzejewski}, \citenamefont {Eckel}, \citenamefont {Murray}, \citenamefont
  {Lanier}, \citenamefont {Edwards}, \citenamefont {Lobb},\ and\ \citenamefont
  {Campbell}}]{Jendrzejewski2014}%
  \BibitemOpen
  \bibfield  {author} {\bibinfo {author} {\bibfnamefont {F.}~\bibnamefont
  {Jendrzejewski}}, \bibinfo {author} {\bibfnamefont {S.}~\bibnamefont
  {Eckel}}, \bibinfo {author} {\bibfnamefont {N.}~\bibnamefont {Murray}},
  \bibinfo {author} {\bibfnamefont {C.}~\bibnamefont {Lanier}}, \bibinfo
  {author} {\bibfnamefont {M.}~\bibnamefont {Edwards}}, \bibinfo {author}
  {\bibfnamefont {C.~J.}\ \bibnamefont {Lobb}},\ and\ \bibinfo {author}
  {\bibfnamefont {G.~K.}\ \bibnamefont {Campbell}},\ }\bibfield  {title}
  {\bibinfo {title} {Resistive flow in a weakly interacting bose-einstein
  condensate},\ }\href {https://doi.org/10.1103/PhysRevLett.113.045305}
  {\bibfield  {journal} {\bibinfo  {journal} {Phys. Rev. Lett.}\ }\textbf
  {\bibinfo {volume} {113}},\ \bibinfo {pages} {045305} (\bibinfo {year}
  {2014})}\BibitemShut {NoStop}%
\bibitem [{\citenamefont {Krinner}\ \emph {et~al.}(2017)\citenamefont
  {Krinner}, \citenamefont {Esslinger},\ and\ \citenamefont
  {Brantut}}]{Krinner2017}%
  \BibitemOpen
  \bibfield  {author} {\bibinfo {author} {\bibfnamefont {S.}~\bibnamefont
  {Krinner}}, \bibinfo {author} {\bibfnamefont {T.}~\bibnamefont {Esslinger}},\
  and\ \bibinfo {author} {\bibfnamefont {J.-P.}\ \bibnamefont {Brantut}},\
  }\bibfield  {title} {\bibinfo {title} {Two-terminal transport measurements
  with cold atoms},\ }\href {https://doi.org/10.1088/1361-648x/aa74a1}
  {\bibfield  {journal} {\bibinfo  {journal} {Journal of Physics: Condensed
  Matter}\ }\textbf {\bibinfo {volume} {29}},\ \bibinfo {pages} {343003}
  (\bibinfo {year} {2017})}\BibitemShut {NoStop}%
\bibitem [{\citenamefont {Gamayun}\ \emph {et~al.}(2021)\citenamefont
  {Gamayun}, \citenamefont {Slobodeniuk}, \citenamefont {Caux},\ and\
  \citenamefont {Lychkovskiy}}]{Gamayun2021}%
  \BibitemOpen
  \bibfield  {author} {\bibinfo {author} {\bibfnamefont {O.}~\bibnamefont
  {Gamayun}}, \bibinfo {author} {\bibfnamefont {A.}~\bibnamefont
  {Slobodeniuk}}, \bibinfo {author} {\bibfnamefont {J.-S.}\ \bibnamefont
  {Caux}},\ and\ \bibinfo {author} {\bibfnamefont {O.}~\bibnamefont
  {Lychkovskiy}},\ }\bibfield  {title} {\bibinfo {title} {Nonequilibrium phase
  transition in transport through a driven quantum point contact},\ }\href
  {https://doi.org/10.1103/PhysRevB.103.L041405} {\bibfield  {journal}
  {\bibinfo  {journal} {Phys. Rev. B}\ }\textbf {\bibinfo {volume} {103}},\
  \bibinfo {pages} {L041405} (\bibinfo {year} {2021})}\BibitemShut {NoStop}%
\bibitem [{\citenamefont {Valtolina}\ \emph {et~al.}(2015)\citenamefont
  {Valtolina}, \citenamefont {Burchianti}, \citenamefont {Amico}, \citenamefont
  {Neri}, \citenamefont {Xhani}, \citenamefont {Seman}, \citenamefont
  {Trombettoni}, \citenamefont {Smerzi}, \citenamefont {Zaccanti},
  \citenamefont {Inguscio},\ and\ \citenamefont {Roati}}]{Valtolina2015}%
  \BibitemOpen
  \bibfield  {author} {\bibinfo {author} {\bibfnamefont {G.}~\bibnamefont
  {Valtolina}}, \bibinfo {author} {\bibfnamefont {A.}~\bibnamefont
  {Burchianti}}, \bibinfo {author} {\bibfnamefont {A.}~\bibnamefont {Amico}},
  \bibinfo {author} {\bibfnamefont {E.}~\bibnamefont {Neri}}, \bibinfo {author}
  {\bibfnamefont {K.}~\bibnamefont {Xhani}}, \bibinfo {author} {\bibfnamefont
  {J.~A.}\ \bibnamefont {Seman}}, \bibinfo {author} {\bibfnamefont
  {A.}~\bibnamefont {Trombettoni}}, \bibinfo {author} {\bibfnamefont
  {A.}~\bibnamefont {Smerzi}}, \bibinfo {author} {\bibfnamefont
  {M.}~\bibnamefont {Zaccanti}}, \bibinfo {author} {\bibfnamefont
  {M.}~\bibnamefont {Inguscio}},\ and\ \bibinfo {author} {\bibfnamefont
  {G.}~\bibnamefont {Roati}},\ }\bibfield  {title} {\bibinfo {title} {Josephson
  effect in fermionic superfluids across the bec-bcs crossover},\ }\href
  {https://doi.org/10.1126/science.aac9725} {\bibfield  {journal} {\bibinfo
  {journal} {Science}\ }\textbf {\bibinfo {volume} {350}},\ \bibinfo {pages}
  {1505} (\bibinfo {year} {2015})}\BibitemShut {NoStop}%
\bibitem [{\citenamefont {Burchianti}\ \emph {et~al.}(2018)\citenamefont
  {Burchianti}, \citenamefont {Scazza}, \citenamefont {Amico}, \citenamefont
  {Valtolina}, \citenamefont {Seman}, \citenamefont {Fort}, \citenamefont
  {Zaccanti}, \citenamefont {Inguscio},\ and\ \citenamefont
  {Roati}}]{Burchianti2018}%
  \BibitemOpen
  \bibfield  {author} {\bibinfo {author} {\bibfnamefont {A.}~\bibnamefont
  {Burchianti}}, \bibinfo {author} {\bibfnamefont {F.}~\bibnamefont {Scazza}},
  \bibinfo {author} {\bibfnamefont {A.}~\bibnamefont {Amico}}, \bibinfo
  {author} {\bibfnamefont {G.}~\bibnamefont {Valtolina}}, \bibinfo {author}
  {\bibfnamefont {J.~A.}\ \bibnamefont {Seman}}, \bibinfo {author}
  {\bibfnamefont {C.}~\bibnamefont {Fort}}, \bibinfo {author} {\bibfnamefont
  {M.}~\bibnamefont {Zaccanti}}, \bibinfo {author} {\bibfnamefont
  {M.}~\bibnamefont {Inguscio}},\ and\ \bibinfo {author} {\bibfnamefont
  {G.}~\bibnamefont {Roati}},\ }\bibfield  {title} {\bibinfo {title}
  {Connecting dissipation and phase slips in a josephson junction between
  fermionic superfluids},\ }\href
  {https://doi.org/10.1103/PhysRevLett.120.025302} {\bibfield  {journal}
  {\bibinfo  {journal} {Phys. Rev. Lett.}\ }\textbf {\bibinfo {volume} {120}},\
  \bibinfo {pages} {025302} (\bibinfo {year} {2018})}\BibitemShut {NoStop}%
\bibitem [{\citenamefont {Zaccanti}\ and\ \citenamefont
  {Zwerger}(2019)}]{Zaccanti2019}%
  \BibitemOpen
  \bibfield  {author} {\bibinfo {author} {\bibfnamefont {M.}~\bibnamefont
  {Zaccanti}}\ and\ \bibinfo {author} {\bibfnamefont {W.}~\bibnamefont
  {Zwerger}},\ }\bibfield  {title} {\bibinfo {title} {Critical josephson
  current in bcs-bec--crossover superfluids},\ }\href
  {https://doi.org/10.1103/PhysRevA.100.063601} {\bibfield  {journal} {\bibinfo
   {journal} {Phys. Rev. A}\ }\textbf {\bibinfo {volume} {100}},\ \bibinfo
  {pages} {063601} (\bibinfo {year} {2019})}\BibitemShut {NoStop}%
\bibitem [{\citenamefont {Luick}\ \emph {et~al.}(2020)\citenamefont {Luick},
  \citenamefont {Sobirey}, \citenamefont {Bohlen}, \citenamefont {Singh},
  \citenamefont {Mathey}, \citenamefont {Lompe},\ and\ \citenamefont
  {Moritz}}]{Luick2020}%
  \BibitemOpen
  \bibfield  {author} {\bibinfo {author} {\bibfnamefont {N.}~\bibnamefont
  {Luick}}, \bibinfo {author} {\bibfnamefont {L.}~\bibnamefont {Sobirey}},
  \bibinfo {author} {\bibfnamefont {M.}~\bibnamefont {Bohlen}}, \bibinfo
  {author} {\bibfnamefont {V.~P.}\ \bibnamefont {Singh}}, \bibinfo {author}
  {\bibfnamefont {L.}~\bibnamefont {Mathey}}, \bibinfo {author} {\bibfnamefont
  {T.}~\bibnamefont {Lompe}},\ and\ \bibinfo {author} {\bibfnamefont
  {H.}~\bibnamefont {Moritz}},\ }\bibfield  {title} {\bibinfo {title} {An ideal
  josephson junction in an ultracold two-dimensional fermi gas},\ }\href
  {https://doi.org/10.1126/science.aaz2342} {\bibfield  {journal} {\bibinfo
  {journal} {Science}\ }\textbf {\bibinfo {volume} {369}},\ \bibinfo {pages}
  {89} (\bibinfo {year} {2020})}\BibitemShut {NoStop}%
\bibitem [{\citenamefont {Xhani}\ \emph
  {et~al.}(2020{\natexlab{a}})\citenamefont {Xhani}, \citenamefont {Neri},
  \citenamefont {Galantucci}, \citenamefont {Scazza}, \citenamefont
  {Burchianti}, \citenamefont {Lee}, \citenamefont {Barenghi}, \citenamefont
  {Trombettoni}, \citenamefont {Inguscio}, \citenamefont {Zaccanti},
  \citenamefont {Roati},\ and\ \citenamefont {Proukakis}}]{Xhani2020a}%
  \BibitemOpen
  \bibfield  {author} {\bibinfo {author} {\bibfnamefont {K.}~\bibnamefont
  {Xhani}}, \bibinfo {author} {\bibfnamefont {E.}~\bibnamefont {Neri}},
  \bibinfo {author} {\bibfnamefont {L.}~\bibnamefont {Galantucci}}, \bibinfo
  {author} {\bibfnamefont {F.}~\bibnamefont {Scazza}}, \bibinfo {author}
  {\bibfnamefont {A.}~\bibnamefont {Burchianti}}, \bibinfo {author}
  {\bibfnamefont {K.-L.}\ \bibnamefont {Lee}}, \bibinfo {author} {\bibfnamefont
  {C.~F.}\ \bibnamefont {Barenghi}}, \bibinfo {author} {\bibfnamefont
  {A.}~\bibnamefont {Trombettoni}}, \bibinfo {author} {\bibfnamefont
  {M.}~\bibnamefont {Inguscio}}, \bibinfo {author} {\bibfnamefont
  {M.}~\bibnamefont {Zaccanti}}, \bibinfo {author} {\bibfnamefont
  {G.}~\bibnamefont {Roati}},\ and\ \bibinfo {author} {\bibfnamefont {N.~P.}\
  \bibnamefont {Proukakis}},\ }\bibfield  {title} {\bibinfo {title} {Critical
  transport and vortex dynamics in a thin atomic josephson junction},\ }\href
  {https://doi.org/10.1103/PhysRevLett.124.045301} {\bibfield  {journal}
  {\bibinfo  {journal} {Phys. Rev. Lett.}\ }\textbf {\bibinfo {volume} {124}},\
  \bibinfo {pages} {045301} (\bibinfo {year} {2020}{\natexlab{a}})}\BibitemShut
  {NoStop}%
\bibitem [{\citenamefont {Kwon}\ \emph {et~al.}(2020)\citenamefont {Kwon},
  \citenamefont {Del~Pace}, \citenamefont {Panza}, \citenamefont {Inguscio},
  \citenamefont {Zwerger}, \citenamefont {Zaccanti}, \citenamefont {Scazza},\
  and\ \citenamefont {Roati}}]{Kwon2020}%
  \BibitemOpen
  \bibfield  {author} {\bibinfo {author} {\bibfnamefont {W.~J.}\ \bibnamefont
  {Kwon}}, \bibinfo {author} {\bibfnamefont {G.}~\bibnamefont {Del~Pace}},
  \bibinfo {author} {\bibfnamefont {R.}~\bibnamefont {Panza}}, \bibinfo
  {author} {\bibfnamefont {M.}~\bibnamefont {Inguscio}}, \bibinfo {author}
  {\bibfnamefont {W.}~\bibnamefont {Zwerger}}, \bibinfo {author} {\bibfnamefont
  {M.}~\bibnamefont {Zaccanti}}, \bibinfo {author} {\bibfnamefont
  {F.}~\bibnamefont {Scazza}},\ and\ \bibinfo {author} {\bibfnamefont
  {G.}~\bibnamefont {Roati}},\ }\bibfield  {title} {\bibinfo {title} {Strongly
  correlated superfluid order parameters from dc josephson supercurrents},\
  }\href {https://doi.org/10.1126/science.aaz2463} {\bibfield  {journal}
  {\bibinfo  {journal} {Science}\ }\textbf {\bibinfo {volume} {369}},\ \bibinfo
  {pages} {84} (\bibinfo {year} {2020})}\BibitemShut {NoStop}%
\bibitem [{\citenamefont {Shin}\ \emph {et~al.}(2004)\citenamefont {Shin},
  \citenamefont {Saba}, \citenamefont {Pasquini}, \citenamefont {Ketterle},
  \citenamefont {Pritchard},\ and\ \citenamefont {Leanhardt}}]{Shin2004}%
  \BibitemOpen
  \bibfield  {author} {\bibinfo {author} {\bibfnamefont {Y.}~\bibnamefont
  {Shin}}, \bibinfo {author} {\bibfnamefont {M.}~\bibnamefont {Saba}}, \bibinfo
  {author} {\bibfnamefont {T.~A.}\ \bibnamefont {Pasquini}}, \bibinfo {author}
  {\bibfnamefont {W.}~\bibnamefont {Ketterle}}, \bibinfo {author}
  {\bibfnamefont {D.~E.}\ \bibnamefont {Pritchard}},\ and\ \bibinfo {author}
  {\bibfnamefont {A.~E.}\ \bibnamefont {Leanhardt}},\ }\bibfield  {title}
  {\bibinfo {title} {Atom interferometry with bose-einstein condensates in a
  double-well potential},\ }\href
  {https://doi.org/10.1103/PhysRevLett.92.050405} {\bibfield  {journal}
  {\bibinfo  {journal} {Phys. Rev. Lett.}\ }\textbf {\bibinfo {volume} {92}},\
  \bibinfo {pages} {050405} (\bibinfo {year} {2004})}\BibitemShut {NoStop}%
\bibitem [{\citenamefont {Schumm}\ \emph {et~al.}(2005)\citenamefont {Schumm},
  \citenamefont {Hofferberth}, \citenamefont {Andersson}, \citenamefont
  {Wildermuth}, \citenamefont {Groth}, \citenamefont {Bar-Joseph},
  \citenamefont {Schmiedmayer},\ and\ \citenamefont {Krüger}}]{Schumm2005}%
  \BibitemOpen
  \bibfield  {author} {\bibinfo {author} {\bibfnamefont {T.}~\bibnamefont
  {Schumm}}, \bibinfo {author} {\bibfnamefont {S.}~\bibnamefont {Hofferberth}},
  \bibinfo {author} {\bibfnamefont {L.~M.}\ \bibnamefont {Andersson}}, \bibinfo
  {author} {\bibfnamefont {S.}~\bibnamefont {Wildermuth}}, \bibinfo {author}
  {\bibfnamefont {S.}~\bibnamefont {Groth}}, \bibinfo {author} {\bibfnamefont
  {I.}~\bibnamefont {Bar-Joseph}}, \bibinfo {author} {\bibfnamefont
  {J.}~\bibnamefont {Schmiedmayer}},\ and\ \bibinfo {author} {\bibfnamefont
  {P.}~\bibnamefont {Krüger}},\ }\bibfield  {title} {\bibinfo {title}
  {Matter-wave interferometry in a double well on an atom chip},\ }\href
  {https://doi.org/10.1038/nphys125} {\bibfield  {journal} {\bibinfo  {journal}
  {Nature Physics}\ }\textbf {\bibinfo {volume} {1}},\ \bibinfo {pages} {57}
  (\bibinfo {year} {2005})}\BibitemShut {NoStop}%
\bibitem [{\citenamefont {Mennemann}\ \emph {et~al.}(2021)\citenamefont
  {Mennemann}, \citenamefont {Mazets}, \citenamefont {Pigneur}, \citenamefont
  {Stimming}, \citenamefont {Mauser}, \citenamefont {Schmiedmayer},\ and\
  \citenamefont {Erne}}]{Mennemann2020}%
  \BibitemOpen
  \bibfield  {author} {\bibinfo {author} {\bibfnamefont {J.-F.}\ \bibnamefont
  {Mennemann}}, \bibinfo {author} {\bibfnamefont {I.~E.}\ \bibnamefont
  {Mazets}}, \bibinfo {author} {\bibfnamefont {M.}~\bibnamefont {Pigneur}},
  \bibinfo {author} {\bibfnamefont {H.~P.}\ \bibnamefont {Stimming}}, \bibinfo
  {author} {\bibfnamefont {N.~J.}\ \bibnamefont {Mauser}}, \bibinfo {author}
  {\bibfnamefont {J.}~\bibnamefont {Schmiedmayer}},\ and\ \bibinfo {author}
  {\bibfnamefont {S.}~\bibnamefont {Erne}},\ }\bibfield  {title} {\bibinfo
  {title} {Relaxation in an extended bosonic josephson junction},\ }\href
  {https://doi.org/10.1103/PhysRevResearch.3.023197} {\bibfield  {journal}
  {\bibinfo  {journal} {Phys. Rev. Research}\ }\textbf {\bibinfo {volume}
  {3}},\ \bibinfo {pages} {023197} (\bibinfo {year} {2021})}\BibitemShut
  {NoStop}%
\bibitem [{\citenamefont {Bidasyuk}\ \emph {et~al.}(2016)\citenamefont
  {Bidasyuk}, \citenamefont {Prikhodko},\ and\ \citenamefont
  {Weyrauch}}]{Bidasyuk2016}%
  \BibitemOpen
  \bibfield  {author} {\bibinfo {author} {\bibfnamefont {Y.~M.}\ \bibnamefont
  {Bidasyuk}}, \bibinfo {author} {\bibfnamefont {O.~O.}\ \bibnamefont
  {Prikhodko}},\ and\ \bibinfo {author} {\bibfnamefont {M.}~\bibnamefont
  {Weyrauch}},\ }\bibfield  {title} {\bibinfo {title} {Phonon-josephson
  resonances in atomtronic circuits},\ }\href
  {https://doi.org/10.1103/PhysRevA.94.033603} {\bibfield  {journal} {\bibinfo
  {journal} {Phys. Rev. A}\ }\textbf {\bibinfo {volume} {94}},\ \bibinfo
  {pages} {033603} (\bibinfo {year} {2016})}\BibitemShut {NoStop}%
\bibitem [{\citenamefont {Gati}\ \emph {et~al.}(2006)\citenamefont {Gati},
  \citenamefont {Hemmerling}, \citenamefont {F\"olling}, \citenamefont
  {Albiez},\ and\ \citenamefont {Oberthaler}}]{Gati2006}%
  \BibitemOpen
  \bibfield  {author} {\bibinfo {author} {\bibfnamefont {R.}~\bibnamefont
  {Gati}}, \bibinfo {author} {\bibfnamefont {B.}~\bibnamefont {Hemmerling}},
  \bibinfo {author} {\bibfnamefont {J.}~\bibnamefont {F\"olling}}, \bibinfo
  {author} {\bibfnamefont {M.}~\bibnamefont {Albiez}},\ and\ \bibinfo {author}
  {\bibfnamefont {M.~K.}\ \bibnamefont {Oberthaler}},\ }\bibfield  {title}
  {\bibinfo {title} {Noise thermometry with two weakly coupled bose-einstein
  condensates},\ }\href {https://doi.org/10.1103/PhysRevLett.96.130404}
  {\bibfield  {journal} {\bibinfo  {journal} {Phys. Rev. Lett.}\ }\textbf
  {\bibinfo {volume} {96}},\ \bibinfo {pages} {130404} (\bibinfo {year}
  {2006})}\BibitemShut {NoStop}%
\bibitem [{\citenamefont {Edwards}(2013)}]{Edwards2013}%
  \BibitemOpen
  \bibfield  {author} {\bibinfo {author} {\bibfnamefont {M.}~\bibnamefont
  {Edwards}},\ }\bibfield  {title} {\bibinfo {title} {Atom squid},\ }\href
  {https://doi.org/10.1038/nphys2546} {\bibfield  {journal} {\bibinfo
  {journal} {Nature Physics}\ }\textbf {\bibinfo {volume} {9}},\ \bibinfo
  {pages} {68} (\bibinfo {year} {2013})}\BibitemShut {NoStop}%
\bibitem [{\citenamefont {Amico}\ \emph {et~al.}(2017)\citenamefont {Amico},
  \citenamefont {Birkl}, \citenamefont {Boshier},\ and\ \citenamefont
  {Kwek}}]{Amico2017}%
  \BibitemOpen
  \bibfield  {author} {\bibinfo {author} {\bibfnamefont {L.}~\bibnamefont
  {Amico}}, \bibinfo {author} {\bibfnamefont {G.}~\bibnamefont {Birkl}},
  \bibinfo {author} {\bibfnamefont {M.}~\bibnamefont {Boshier}},\ and\ \bibinfo
  {author} {\bibfnamefont {L.-C.}\ \bibnamefont {Kwek}},\ }\bibfield  {title}
  {\bibinfo {title} {{Focus on atomtronics-enabled quantum technologies}},\
  }\href {https://doi.org/10.1088/1367-2630/aa5a6d} {\bibfield  {journal}
  {\bibinfo  {journal} {New Journal of Physics}\ }\textbf {\bibinfo {volume}
  {19}},\ \bibinfo {pages} {020201} (\bibinfo {year} {2017})}\BibitemShut
  {NoStop}%
\bibitem [{\citenamefont {Ryu}\ \emph {et~al.}(2020)\citenamefont {Ryu},
  \citenamefont {Samson},\ and\ \citenamefont {Boshier}}]{Ryu2020}%
  \BibitemOpen
  \bibfield  {author} {\bibinfo {author} {\bibfnamefont {C.}~\bibnamefont
  {Ryu}}, \bibinfo {author} {\bibfnamefont {E.~C.}\ \bibnamefont {Samson}},\
  and\ \bibinfo {author} {\bibfnamefont {M.~G.}\ \bibnamefont {Boshier}},\
  }\bibfield  {title} {\bibinfo {title} {Quantum interference of currents in an
  atomtronic squid},\ }\href {https://doi.org/10.1038/s41467-020-17185-6}
  {\bibfield  {journal} {\bibinfo  {journal} {Nature Communications}\ }\textbf
  {\bibinfo {volume} {11}},\ \bibinfo {pages} {3338} (\bibinfo {year}
  {2020})}\BibitemShut {NoStop}%
\bibitem [{\citenamefont {Abad}\ \emph {et~al.}(2015)\citenamefont {Abad},
  \citenamefont {Guilleumas}, \citenamefont {Mayol}, \citenamefont {Piazza},
  \citenamefont {Jezek},\ and\ \citenamefont {Smerzi}}]{Abad2015}%
  \BibitemOpen
  \bibfield  {author} {\bibinfo {author} {\bibfnamefont {M.}~\bibnamefont
  {Abad}}, \bibinfo {author} {\bibfnamefont {M.}~\bibnamefont {Guilleumas}},
  \bibinfo {author} {\bibfnamefont {R.}~\bibnamefont {Mayol}}, \bibinfo
  {author} {\bibfnamefont {F.}~\bibnamefont {Piazza}}, \bibinfo {author}
  {\bibfnamefont {D.~M.}\ \bibnamefont {Jezek}},\ and\ \bibinfo {author}
  {\bibfnamefont {A.}~\bibnamefont {Smerzi}},\ }\bibfield  {title} {\bibinfo
  {title} {{Phase slips and vortex dynamics in Josephson oscillations between
  Bose-Einstein condensates}},\ }\href
  {https://doi.org/10.1209/0295-5075/109/40005} {\bibfield  {journal} {\bibinfo
   {journal} {EPL (Europhysics Letters)}\ }\textbf {\bibinfo {volume} {109}},\
  \bibinfo {pages} {40005} (\bibinfo {year} {2015})},\ \Eprint
  {https://arxiv.org/abs/1409.5598} {1409.5598} \BibitemShut {NoStop}%
\bibitem [{\citenamefont {Saha}\ \emph {et~al.}(2019)\citenamefont {Saha},
  \citenamefont {Adhikary}, \citenamefont {Mal}, \citenamefont {Dastidar},\
  and\ \citenamefont {Deb}}]{Saha2019}%
  \BibitemOpen
  \bibfield  {author} {\bibinfo {author} {\bibfnamefont {A.~K.}\ \bibnamefont
  {Saha}}, \bibinfo {author} {\bibfnamefont {K.}~\bibnamefont {Adhikary}},
  \bibinfo {author} {\bibfnamefont {S.}~\bibnamefont {Mal}}, \bibinfo {author}
  {\bibfnamefont {K.~R.}\ \bibnamefont {Dastidar}},\ and\ \bibinfo {author}
  {\bibfnamefont {B.}~\bibnamefont {Deb}},\ }\bibfield  {title} {\bibinfo
  {title} {{The effects of trap-confinement and interatomic interactions on
  Josephson effects and macroscopic quantum self-trapping for a Bose-Einstein
  condensate}},\ }\bibfield  {journal} {\bibinfo  {journal} {Journal of Physics
  B: Atomic, Molecular and Optical Physics}\ }\textbf {\bibinfo {volume}
  {52}},\ \href {https://doi.org/10.1088/1361-6455/ab2b58}
  {10.1088/1361-6455/ab2b58} (\bibinfo {year} {2019})\BibitemShut {NoStop}%
\bibitem [{\citenamefont {Piazza}\ \emph {et~al.}(2011)\citenamefont {Piazza},
  \citenamefont {Collins},\ and\ \citenamefont {Smerzi}}]{Piazza2011}%
  \BibitemOpen
  \bibfield  {author} {\bibinfo {author} {\bibfnamefont {F.}~\bibnamefont
  {Piazza}}, \bibinfo {author} {\bibfnamefont {L.~a.}\ \bibnamefont
  {Collins}},\ and\ \bibinfo {author} {\bibfnamefont {A.}~\bibnamefont
  {Smerzi}},\ }\bibfield  {title} {\bibinfo {title} {{Instability and vortex
  ring dynamics in a three-dimensional superfluid flow through a
  constriction}},\ }\href {https://doi.org/10.1088/1367-2630/13/4/043008}
  {\bibfield  {journal} {\bibinfo  {journal} {New Journal of Physics}\ }\textbf
  {\bibinfo {volume} {13}},\ \bibinfo {pages} {043008} (\bibinfo {year}
  {2011})}\BibitemShut {NoStop}%
\bibitem [{\citenamefont {Singh}\ \emph {et~al.}(2020)\citenamefont {Singh},
  \citenamefont {Luick}, \citenamefont {Sobirey},\ and\ \citenamefont
  {Mathey}}]{Singh2020}%
  \BibitemOpen
  \bibfield  {author} {\bibinfo {author} {\bibfnamefont {V.~P.}\ \bibnamefont
  {Singh}}, \bibinfo {author} {\bibfnamefont {N.}~\bibnamefont {Luick}},
  \bibinfo {author} {\bibfnamefont {L.}~\bibnamefont {Sobirey}},\ and\ \bibinfo
  {author} {\bibfnamefont {L.}~\bibnamefont {Mathey}},\ }\bibfield  {title}
  {\bibinfo {title} {Josephson junction dynamics in a two-dimensional ultracold
  bose gas},\ }\href {https://doi.org/10.1103/PhysRevResearch.2.033298}
  {\bibfield  {journal} {\bibinfo  {journal} {Phys. Rev. Research}\ }\textbf
  {\bibinfo {volume} {2}},\ \bibinfo {pages} {033298} (\bibinfo {year}
  {2020})}\BibitemShut {NoStop}%
\bibitem [{\citenamefont {Griffin}\ \emph {et~al.}(2020)\citenamefont
  {Griffin}, \citenamefont {Nazarenko},\ and\ \citenamefont
  {Proment}}]{Griffin2020a}%
  \BibitemOpen
  \bibfield  {author} {\bibinfo {author} {\bibfnamefont {A.}~\bibnamefont
  {Griffin}}, \bibinfo {author} {\bibfnamefont {S.}~\bibnamefont {Nazarenko}},\
  and\ \bibinfo {author} {\bibfnamefont {D.}~\bibnamefont {Proment}},\
  }\bibfield  {title} {\bibinfo {title} {{Breaking of Josephson junction
  oscillations and onset of quantum turbulence in Bose–Einstein
  condensates}},\ }\href {https://doi.org/10.1088/1751-8121/ab7ad0} {\bibfield
  {journal} {\bibinfo  {journal} {Journal of Physics A: Mathematical and
  Theoretical}\ }\textbf {\bibinfo {volume} {53}},\ \bibinfo {pages} {175701}
  (\bibinfo {year} {2020})}\BibitemShut {NoStop}%
\bibitem [{\citenamefont {Simmons}\ \emph {et~al.}(2020)\citenamefont
  {Simmons}, \citenamefont {Bayocboc}, \citenamefont {Pillay}, \citenamefont
  {Colas}, \citenamefont {McCulloch},\ and\ \citenamefont
  {Kheruntsyan}}]{Simmons2020}%
  \BibitemOpen
  \bibfield  {author} {\bibinfo {author} {\bibfnamefont {S.~A.}\ \bibnamefont
  {Simmons}}, \bibinfo {author} {\bibfnamefont {F.~A.}\ \bibnamefont
  {Bayocboc}}, \bibinfo {author} {\bibfnamefont {J.~C.}\ \bibnamefont
  {Pillay}}, \bibinfo {author} {\bibfnamefont {D.}~\bibnamefont {Colas}},
  \bibinfo {author} {\bibfnamefont {I.~P.}\ \bibnamefont {McCulloch}},\ and\
  \bibinfo {author} {\bibfnamefont {K.~V.}\ \bibnamefont {Kheruntsyan}},\
  }\bibfield  {title} {\bibinfo {title} {What is a quantum shock wave?},\
  }\href {https://doi.org/10.1103/PhysRevLett.125.180401} {\bibfield  {journal}
  {\bibinfo  {journal} {Phys. Rev. Lett.}\ }\textbf {\bibinfo {volume} {125}},\
  \bibinfo {pages} {180401} (\bibinfo {year} {2020})}\BibitemShut {NoStop}%
\bibitem [{\citenamefont {Dubessy}\ \emph {et~al.}(2021)\citenamefont
  {Dubessy}, \citenamefont {Polo}, \citenamefont {Perrin}, \citenamefont
  {Minguzzi},\ and\ \citenamefont {Olshanii}}]{Dubessy2021}%
  \BibitemOpen
  \bibfield  {author} {\bibinfo {author} {\bibfnamefont {R.}~\bibnamefont
  {Dubessy}}, \bibinfo {author} {\bibfnamefont {J.}~\bibnamefont {Polo}},
  \bibinfo {author} {\bibfnamefont {H.}~\bibnamefont {Perrin}}, \bibinfo
  {author} {\bibfnamefont {A.}~\bibnamefont {Minguzzi}},\ and\ \bibinfo
  {author} {\bibfnamefont {M.}~\bibnamefont {Olshanii}},\ }\bibfield  {title}
  {\bibinfo {title} {Universal shock-wave propagation in one-dimensional bose
  fluids},\ }\href {https://doi.org/10.1103/PhysRevResearch.3.013098}
  {\bibfield  {journal} {\bibinfo  {journal} {Phys. Rev. Research}\ }\textbf
  {\bibinfo {volume} {3}},\ \bibinfo {pages} {013098} (\bibinfo {year}
  {2021})}\BibitemShut {NoStop}%
\bibitem [{\citenamefont {Mewes}\ \emph {et~al.}(1996)\citenamefont {Mewes},
  \citenamefont {Andrews}, \citenamefont {{Van Druten}}, \citenamefont {Kurn},
  \citenamefont {Durfee},\ and\ \citenamefont {Ketterle}}]{Mewes1996}%
  \BibitemOpen
  \bibfield  {author} {\bibinfo {author} {\bibfnamefont {M.~O.}\ \bibnamefont
  {Mewes}}, \bibinfo {author} {\bibfnamefont {M.~R.}\ \bibnamefont {Andrews}},
  \bibinfo {author} {\bibfnamefont {N.~J.}\ \bibnamefont {{Van Druten}}},
  \bibinfo {author} {\bibfnamefont {D.~M.}\ \bibnamefont {Kurn}}, \bibinfo
  {author} {\bibfnamefont {D.~S.}\ \bibnamefont {Durfee}},\ and\ \bibinfo
  {author} {\bibfnamefont {W.}~\bibnamefont {Ketterle}},\ }\bibfield  {title}
  {\bibinfo {title} {{Bose-einstein condensation in a tightly confining dc
  magnetic trap}},\ }\href {https://doi.org/10.1103/PhysRevLett.77.416}
  {\bibfield  {journal} {\bibinfo  {journal} {Physical Review Letters}\
  }\textbf {\bibinfo {volume} {77}},\ \bibinfo {pages} {416} (\bibinfo {year}
  {1996})}\BibitemShut {NoStop}%
\bibitem [{\citenamefont {G{\"{o}}rlitz}\ \emph {et~al.}(2001)\citenamefont
  {G{\"{o}}rlitz}, \citenamefont {Vogels}, \citenamefont {Leanhardt},
  \citenamefont {Raman}, \citenamefont {Gustavson}, \citenamefont {Abo-Shaeer},
  \citenamefont {Chikkatur}, \citenamefont {Gupta}, \citenamefont {Inouye},
  \citenamefont {Rosenband},\ and\ \citenamefont {Ketterle}}]{Gorlitz2001}%
  \BibitemOpen
  \bibfield  {author} {\bibinfo {author} {\bibfnamefont {A.}~\bibnamefont
  {G{\"{o}}rlitz}}, \bibinfo {author} {\bibfnamefont {J.~M.}\ \bibnamefont
  {Vogels}}, \bibinfo {author} {\bibfnamefont {A.~E.}\ \bibnamefont
  {Leanhardt}}, \bibinfo {author} {\bibfnamefont {C.}~\bibnamefont {Raman}},
  \bibinfo {author} {\bibfnamefont {T.~L.}\ \bibnamefont {Gustavson}}, \bibinfo
  {author} {\bibfnamefont {J.~R.}\ \bibnamefont {Abo-Shaeer}}, \bibinfo
  {author} {\bibfnamefont {A.~P.}\ \bibnamefont {Chikkatur}}, \bibinfo {author}
  {\bibfnamefont {S.}~\bibnamefont {Gupta}}, \bibinfo {author} {\bibfnamefont
  {S.}~\bibnamefont {Inouye}}, \bibinfo {author} {\bibfnamefont
  {T.}~\bibnamefont {Rosenband}},\ and\ \bibinfo {author} {\bibfnamefont
  {W.}~\bibnamefont {Ketterle}},\ }\bibfield  {title} {\bibinfo {title}
  {{Realization of Bose-Einstein condensates in lower dimensions.}},\ }\href
  {https://doi.org/10.1103/PhysRevLett.87.130402} {\bibfield  {journal}
  {\bibinfo  {journal} {Physical review letters}\ }\textbf {\bibinfo {volume}
  {87}},\ \bibinfo {pages} {130402} (\bibinfo {year} {2001})}\BibitemShut
  {NoStop}%
\bibitem [{\citenamefont {Salasnich}\ \emph {et~al.}(2004)\citenamefont
  {Salasnich}, \citenamefont {Parola},\ and\ \citenamefont
  {Reatto}}]{Salasnich2004}%
  \BibitemOpen
  \bibfield  {author} {\bibinfo {author} {\bibfnamefont {L.}~\bibnamefont
  {Salasnich}}, \bibinfo {author} {\bibfnamefont {A.}~\bibnamefont {Parola}},\
  and\ \bibinfo {author} {\bibfnamefont {L.}~\bibnamefont {Reatto}},\
  }\bibfield  {title} {\bibinfo {title} {{Dimensional reduction in
  Bose-Einstein-condensed alkali-metal vapors}},\ }\href
  {https://doi.org/10.1103/PhysRevA.69.045601} {\bibfield  {journal} {\bibinfo
  {journal} {Physical Review A}\ }\textbf {\bibinfo {volume} {69}},\ \bibinfo
  {pages} {045601} (\bibinfo {year} {2004})}\BibitemShut {NoStop}%
\bibitem [{\citenamefont {Olshanii}(1998)}]{Olshanii1998}%
  \BibitemOpen
  \bibfield  {author} {\bibinfo {author} {\bibfnamefont {M.}~\bibnamefont
  {Olshanii}},\ }\bibfield  {title} {\bibinfo {title} {Atomic scattering in the
  presence of an external confinement and a gas of impenetrable bosons},\
  }\href {https://doi.org/10.1103/PhysRevLett.81.938} {\bibfield  {journal}
  {\bibinfo  {journal} {Phys. Rev. Lett.}\ }\textbf {\bibinfo {volume} {81}},\
  \bibinfo {pages} {938} (\bibinfo {year} {1998})}\BibitemShut {NoStop}%
\bibitem [{\citenamefont {Blakie}(2008)}]{Blakie2008}%
  \BibitemOpen
  \bibfield  {author} {\bibinfo {author} {\bibfnamefont {P.~B.}\ \bibnamefont
  {Blakie}},\ }\bibfield  {title} {\bibinfo {title} {Numerical method for
  evolving the projected gross-pitaevskii equation},\ }\href
  {https://doi.org/10.1103/PhysRevE.78.026704} {\bibfield  {journal} {\bibinfo
  {journal} {Phys. Rev. E}\ }\textbf {\bibinfo {volume} {78}},\ \bibinfo
  {pages} {026704} (\bibinfo {year} {2008})}\BibitemShut {NoStop}%
\bibitem [{Note1()}]{Note1}%
  \BibitemOpen
  \bibinfo {note} {We use MATLAB ode45 integrator and have checked that
  relative errors on total atom number and energy remain in the $10^{-10}$ to
  $10^{-6}$ range for all simulations.}\BibitemShut {Stop}%
\bibitem [{\citenamefont {Antoine}\ \emph {et~al.}(2017)\citenamefont
  {Antoine}, \citenamefont {Besse}, \citenamefont {Duboscq},\ and\
  \citenamefont {Rispoli}}]{Antoine2007}%
  \BibitemOpen
  \bibfield  {author} {\bibinfo {author} {\bibfnamefont {X.}~\bibnamefont
  {Antoine}}, \bibinfo {author} {\bibfnamefont {C.}~\bibnamefont {Besse}},
  \bibinfo {author} {\bibfnamefont {R.}~\bibnamefont {Duboscq}},\ and\ \bibinfo
  {author} {\bibfnamefont {V.}~\bibnamefont {Rispoli}},\ }\bibfield  {title}
  {\bibinfo {title} {Acceleration of the imaginary time method for spectrally
  computing the stationary states of gross–pitaevskii equations},\ }\href
  {https://doi.org/https://doi.org/10.1016/j.cpc.2017.05.008} {\bibfield
  {journal} {\bibinfo  {journal} {Computer Physics Communications}\ }\textbf
  {\bibinfo {volume} {219}},\ \bibinfo {pages} {70} (\bibinfo {year}
  {2017})}\BibitemShut {NoStop}%
\bibitem [{\citenamefont {Barenghi}\ and\ \citenamefont
  {Parker}(2016)}]{Barenghi2016}%
  \BibitemOpen
  \bibfield  {author} {\bibinfo {author} {\bibfnamefont {C.~F.}\ \bibnamefont
  {Barenghi}}\ and\ \bibinfo {author} {\bibfnamefont {N.~G.}\ \bibnamefont
  {Parker}},\ }\href@noop {} {\emph {\bibinfo {title} {A Primer on Quantum
  Fluids}}}\ (\bibinfo  {publisher} {Springer},\ \bibinfo {address} {Berlin},\
  \bibinfo {year} {2016})\BibitemShut {NoStop}%
\bibitem [{Note2()}]{Note2}%
  \BibitemOpen
  \bibinfo {note} {During imaginary time propagation we monitor the chemical
  potential at each step and stop when relative changes are below the
  $10^{-12}$ level.}\BibitemShut {Stop}%
\bibitem [{Note3()}]{Note3}%
  \BibitemOpen
  \bibinfo {note} {We use here the expectation value of the momentum
  operator.}\BibitemShut {Stop}%
\bibitem [{\citenamefont {Xhani}\ \emph
  {et~al.}(2020{\natexlab{b}})\citenamefont {Xhani}, \citenamefont
  {Galantucci}, \citenamefont {Barenghi}, \citenamefont {Roati}, \citenamefont
  {Trombettoni},\ and\ \citenamefont {Proukakis}}]{Xhani2020}%
  \BibitemOpen
  \bibfield  {author} {\bibinfo {author} {\bibfnamefont {K.}~\bibnamefont
  {Xhani}}, \bibinfo {author} {\bibfnamefont {L.}~\bibnamefont {Galantucci}},
  \bibinfo {author} {\bibfnamefont {C.~F.}\ \bibnamefont {Barenghi}}, \bibinfo
  {author} {\bibfnamefont {G.}~\bibnamefont {Roati}}, \bibinfo {author}
  {\bibfnamefont {A.}~\bibnamefont {Trombettoni}},\ and\ \bibinfo {author}
  {\bibfnamefont {N.~P.}\ \bibnamefont {Proukakis}},\ }\bibfield  {title}
  {\bibinfo {title} {Dynamical phase diagram of ultracold josephson
  junctions},\ }\href {https://doi.org/10.1088/1367-2630/abc8e4} {\bibfield
  {journal} {\bibinfo  {journal} {New Journal of Physics}\ }\textbf {\bibinfo
  {volume} {22}},\ \bibinfo {pages} {123006} (\bibinfo {year}
  {2020}{\natexlab{b}})}\BibitemShut {NoStop}%
\bibitem [{\citenamefont {El}\ and\ \citenamefont {Hoefer}(2016)}]{El2016a}%
  \BibitemOpen
  \bibfield  {author} {\bibinfo {author} {\bibfnamefont {G.}~\bibnamefont
  {El}}\ and\ \bibinfo {author} {\bibfnamefont {M.}~\bibnamefont {Hoefer}},\
  }\bibfield  {title} {\bibinfo {title} {{Dispersive shock waves and modulation
  theory}},\ }\href {https://doi.org/10.1016/j.physd.2016.04.006} {\bibfield
  {journal} {\bibinfo  {journal} {Physica D: Nonlinear Phenomena}\ }\textbf
  {\bibinfo {volume} {333}},\ \bibinfo {pages} {11} (\bibinfo {year}
  {2016})}\BibitemShut {NoStop}%
\bibitem [{\citenamefont {Hakim}(1997)}]{Hakim1997}%
  \BibitemOpen
  \bibfield  {author} {\bibinfo {author} {\bibfnamefont {V.}~\bibnamefont
  {Hakim}},\ }\bibfield  {title} {\bibinfo {title} {Nonlinear schr\"odinger
  flow past an obstacle in one dimension},\ }\href
  {https://doi.org/10.1103/PhysRevE.55.2835} {\bibfield  {journal} {\bibinfo
  {journal} {Phys. Rev. E}\ }\textbf {\bibinfo {volume} {55}},\ \bibinfo
  {pages} {2835} (\bibinfo {year} {1997})}\BibitemShut {NoStop}%
\bibitem [{\citenamefont {Tinkham}(1996)}]{Tinkham1996}%
  \BibitemOpen
  \bibfield  {author} {\bibinfo {author} {\bibfnamefont {M.}~\bibnamefont
  {Tinkham}},\ }\href@noop {} {\emph {\bibinfo {title} {Introduction to
  Superconductivity, 2nd Ed.}}}\ (\bibinfo  {publisher} {McGraw-Hill},\
  \bibinfo {address} {New York},\ \bibinfo {year} {1996})\BibitemShut {NoStop}%
\bibitem [{\citenamefont {Tajik}\ \emph {et~al.}(2019)\citenamefont {Tajik},
  \citenamefont {Rauer}, \citenamefont {Schweigler}, \citenamefont {Cataldini},
  \citenamefont {{a}o Sabino}, \citenamefont {M{\o}ller}, \citenamefont {Ji},
  \citenamefont {Mazets},\ and\ \citenamefont
  {Schmiedmayer}}]{Mohammadamin2019}%
  \BibitemOpen
  \bibfield  {author} {\bibinfo {author} {\bibfnamefont {M.}~\bibnamefont
  {Tajik}}, \bibinfo {author} {\bibfnamefont {B.}~\bibnamefont {Rauer}},
  \bibinfo {author} {\bibfnamefont {T.}~\bibnamefont {Schweigler}}, \bibinfo
  {author} {\bibfnamefont {F.}~\bibnamefont {Cataldini}}, \bibinfo {author}
  {\bibfnamefont {J.}~\bibnamefont {{a}o Sabino}}, \bibinfo {author}
  {\bibfnamefont {F.~S.}\ \bibnamefont {M{\o}ller}}, \bibinfo {author}
  {\bibfnamefont {S.-C.}\ \bibnamefont {Ji}}, \bibinfo {author} {\bibfnamefont
  {I.~E.}\ \bibnamefont {Mazets}},\ and\ \bibinfo {author} {\bibfnamefont
  {J.}~\bibnamefont {Schmiedmayer}},\ }\bibfield  {title} {\bibinfo {title}
  {Designing arbitrary one-dimensional potentials on an atom chip},\ }\href
  {https://doi.org/10.1364/OE.27.033474} {\bibfield  {journal} {\bibinfo
  {journal} {Opt. Express}\ }\textbf {\bibinfo {volume} {27}},\ \bibinfo
  {pages} {33474} (\bibinfo {year} {2019})}\BibitemShut {NoStop}%
\bibitem [{\citenamefont {Xu}\ \emph {et~al.}(2017)\citenamefont {Xu},
  \citenamefont {Conforti}, \citenamefont {Kudlinski}, \citenamefont {Mussot},\
  and\ \citenamefont {Trillo}}]{Xu2017}%
  \BibitemOpen
  \bibfield  {author} {\bibinfo {author} {\bibfnamefont {G.}~\bibnamefont
  {Xu}}, \bibinfo {author} {\bibfnamefont {M.}~\bibnamefont {Conforti}},
  \bibinfo {author} {\bibfnamefont {A.}~\bibnamefont {Kudlinski}}, \bibinfo
  {author} {\bibfnamefont {A.}~\bibnamefont {Mussot}},\ and\ \bibinfo {author}
  {\bibfnamefont {S.}~\bibnamefont {Trillo}},\ }\bibfield  {title} {\bibinfo
  {title} {Dispersive dam-break flow of a photon fluid},\ }\href
  {https://doi.org/10.1103/PhysRevLett.118.254101} {\bibfield  {journal}
  {\bibinfo  {journal} {Phys. Rev. Lett.}\ }\textbf {\bibinfo {volume} {118}},\
  \bibinfo {pages} {254101} (\bibinfo {year} {2017})}\BibitemShut {NoStop}%
\bibitem [{Note4()}]{Note4}%
  \BibitemOpen
  \bibinfo {note} {In particular we neglect the anomalous terms involving the
  fields complex conjugates, as is usual in such expansions.}\BibitemShut
  {Stop}%
\bibitem [{Note5()}]{Note5}%
  \BibitemOpen
  \bibinfo {note} {This constrain is naturally implemented with a spectral
  scheme using the discrete cosine (sine) transform to find even (odd)
  states.}\BibitemShut {Stop}%
\end{thebibliography}%
%

\end{document}